# Frequency-domain "single-shot" (FDSS) ultrafast transient absorption spectroscopy using compressed laser pulses. Part I. Basic treatment. [1]

Ilya A. Shkrob[*],[a] Dmitri A. Oulianov,[a] Robert A. Crowell,[a] and Stanislas Pommeret [a,b]

[a] *Chemistry Division , Argonne National Laboratory,  Argonne, IL 60439*
[b] *CEA/Saclay, DSM/DRECAM/SCM/URA 331 CNRS 91191 Gif-Sur-Yvette Cedex, France*

**Abstract**

Single-shot ultrafast absorbance spectroscopy based on the frequency encoding of the kinetics is analyzed theoretically and implemented experimentally. The kinetics are sampled in the frequency domain using linearly chirped, amplified 33 fs FWHM pulses derived from a Ti:sapphire laser. A variable length grating pair compressor is used to achieve the time resolution of 500-1000 channels per a 2-to-160 ps window with sensitivity $> 5 \times 10^{-4}$. In terms of the acquisition time, FDSS has an advantage over the pump-probe spectroscopy in a situation when the "noise" is dominated by amplitude variations of the signal, due to the pump and flow instabilities. The possibilities of FDSS are illustrated with the kinetics obtained in multiphoton ionization of water and aqueous iodide and one-photon excitation of polycrystalline ZnSe and thin-film amorphous Si:H. Unlike other "single-shot" techniques, FDSS can be implemented for fluid samples flowing in a high-speed jet and for thin solid samples that exhibit interference fringes; no a priori knowledge of the excitation profile of the pump across the beam is needed. Another advantage is that due to the interference of quasimonochromatic components of the chirped probe pulse, an oscillation pattern near the origin of the FDSS kinetics emerges. This pattern is unique and can be used to determine the complex dielectric function of the photogenerated species.

PACS numbers: 42.65.Re, 42.65.-k, 42.62.Fi, 42.25.Hz, 41.75.Lx

______________________________________________________________

* To whom correspondence should be addressed: *Tel* 630-2529516, *FAX* 630-2524993, *e-mail:* shkrob@anl.gov.

# I. INTRODUCTION

Most ultrafast laser spectroscopy experiments are done in a pump-probe fashion: a short probe pulse is delayed in time and then crossed in the sample with a pump pulse. The kinetics are obtained by plotting the transmission/reflection of the probe pulse as a function of the delay time. If the amplitude of the pump-induced kinetics varies from shot to shot, this variation translates into noise that is spread across the whole trace. For example, in multiphoton ionization, small fluctuations of the pump intensity cause large variations in the photoinduced signal. This "noise" may be large even if the transient absorption (TA) itself is strong. Other common sources of the "noise" include flow instabilities, thermal lensing, and photodegradation of the sample. The problem is further exacerbated when one is interested in small variations of a TA signal caused by an external perturbation: an electric or magnetic field, a change in the polarization, a second (delayed) pump pulse of a different energy, etc. If this perturbation causes a change in the TA that is smaller than a typical shot-to-shot variation of the signal amplitude, extensive averaging is needed to extract the data. Last but not least, since the translation stage (used to delay the pulses in time) is a mechanical device, it is not uncommon that a large fraction of the acquisition time, especially in the sub-ns regime, is spent on moving the stage rather than sampling the data. Most of these problems could be solved if there were a reliable method for "single-shot" sampling of the kinetics. By that we imply that the entire kinetic trace is sampled for each probe pulse; still, one may need to average over many such traces to obtain an acceptable S/N ratio. Since the entire kinetics are modulated in the same way, shot-to-shot variations do not produce "noise" and the averaging efficiency is improved.

Our interest in "single-shot" methodology was instigated by the development of an ultrafast high energy electron source for radiation chemistry studies at Argonne. Short (0.5-2 ps) pulses of energetic (1-20 MeV) electrons are ejected from a supersonic helium jet irradiated by >$10^{18}$ W/cm$^2$, 25-35 fs FWHM light pulses. Due to the high power load on the amplification optics, the repetition rate is low, 5-10 Hz (which is already a dramatic improvement on < 1 mHz rate achieved elsewhere). [1] Owing to large shot-to-shot variations in the yield of electrons and the low repetition rate, pump-probe methodology is unsuited for this kind of pump source.

Two approaches for "single-shot" ultrafast spectroscopy have been previously suggested: (i) spatial encoding and (ii) frequency encoding. The former method was proposed in 1970 by Rentzepis and co-workers [2,3] and further developed by Nelson and co-workers. [4,5,6] The method is similar in principle to the one used in frequency-resolved optical gating (FROG) devices. [7] In a typical setup, [4,5] cylindrically focussed pump and probe beams are crossed in the sample, and the transmitted probe and reference beams are imaged on a CCD. Since the arrival time of the probe pulse at the sample continuously changes across the beam (on a sub-ps time scale), the pump and probe collide at different times and this "imprints" the kinetics on the axial profile of transmitted light. The spread of the delay times can be increased further out to 10-15 ps by passing the probe (and/or the pump) beam through an echelon [3,6] or a supersonic beam. [8] Transient absorption, [2,4,5,6] stimulated emission, [3] and grating [5] measurements on picosecond (ps) [2,3] and femtosecond (fs) [4,5] time scales have been carried out in this manner. A more sophisticated version of the same technique employs two crossed echelons to generate a 2D train of probe pulses separated by 30 fs in time that extends out to 10 ps; a CCD is used to analyze the resulting 2D pattern. [6] While spatial encoding may work for some applications, there are several drawbacks that make this technique unsuitable for time resolved radiation chemistry:

First, it is almost impossible to extend the time window of the observation beyond the first 10-15 ps. Second, refraction of the probe and pump beams, by liquid jets and thermal lenses, is a serious concern, as it destroys the pattern. For thin films, interference of the probe beam in the sample ruins the spatio-temporal correlation, and there are no examples of using the technique for samples thinner than 100 μm. Third, to obtain the spatial profile of the probe beam, one needs to know the spatial profile of photogenerated species, [4,5] as it is the product of these two profiles that is registered by the CCD. To this end, a separate experiment with a calibrated sample is needed to obtain the pump beam profile. [4] However, if the photon order of the excitation is not known beforehand or the excitation is mixed order, as often is the case, the knowledge of such a profile would be of little use. Finally, in radiation chemistry applications, it is impossible to focus the ultrashort pulse of relativistic electrons cylindrically or delay this pulse in time without substantial increase in its pulse duration, due to the energy spread. From the engineering prospective, spatio-temporal encoding requires extremely high quality of the light beams and close placement of the detector to the sample, which makes it almost impossible to implement this method in the confined space of a vacuum chamber.

The second technique, frequency encoding, is based on introducing linear chirp (frequency-dependent phase) in the probe pulse. [9,10] As a result, the blue light component of the probe pulse arrives at the sample at a different delay time than the red light component, and the kinetics is imprinted in the *spectrum* of the transmitted probe pulse, with the "delay time" equal to the group delay at a given frequency. [10] Frequency encoding has been used for "single-shot" temporal and spatio-temporal characterization of short light [11,12] and free electron [13] pulses, correlation spectroscopy, [14] holography, [15a] interferometry, [16,17] electro-optic measurements of THz pulses, [15b] etc. Furthermore, this encoding is implicit in the pump-probe spectroscopies that use chirped laser pulses, including several variants of transient absorption, [18,19] reflection, [19] interferometry, [17] and 4-wave mixing [20]. Nevertheless, we are aware of only one previous work [10] in which frequency encoding was used for "single-shot" *frequency domain* detection of the TA kinetics. In 1992, Beddard et al. [10] selected a 600-650 nm pulse from white light supercontinuum and stretched it to 8-20 ps using a 0.5-2.5 m single mode optical fiber. This chirped probe pulse was used to study the bleaching of a malachite green dye in an alcohol solution. TA kinetics that exhibited the S/N ratio of 100:1 for a $10^{-2}$ bleaching signal were obtained. When the 625 nm pump pulse was shortened from 500 fs to 250 fs FWHM, an oscillatory pattern superimposed on the kinetics emerged. Beddard et al. [10] interpreted these oscillations as nonlinear interference effects; in fact, as shown in section 2, these oscillations are the inherent feature of the frequency-domain "single-shot" (FDSS) spectroscopy. Rather than being detrimental, these oscillation carry the information on the phase of the complex dielectric function (similar to the way the chirped-pulse nonlinear refraction spectroscopy does) [19] which thereby can be retrieved.

In the decade that passed after the initial experiments of Beddard et al.,[10] generation of ultrashort pulses using chirped-pulse amplification (CPA) with grating stretchers and compressors became routine.[9] These days, there is no need to use divergent, low-quality white light supercontinuum pulses for FDSS since femtosecond pulses derived from an amplified Kerr-lens mode-locked Ti:sapphire laser already provide the bandwidth that is optimum for FDSS measurements. Due to their high quality, these fs pulses can be negatively chirped using grating compressors that provide group velocity dispersion up to a few ps$^2$.[9] As argued in section III.1, with the typical compressor designs that are used for CPA, "single-shot" kinetic measurements over 300-500 ps are possible.

With these developments, what was exotic in 1992 has a potential to become a mainstream technique. In this work, we showcase FDSS for several photosystems, including a 2 mm thick polycrystalline ZnSe sample (section V.1), a 1.4 μm thin film of amorphous Si:H alloy (Appendix 2), aqueous sodium iodide (flowed in a glass cell and a high-speed jet) from which a hydrated electron is detached after absorption of two 400 nm photons (section V.2), and liquid water (flowed in a 150 μm thick high-speed jet) ionized by absorption of three 400 nm photons (section V.3). It is shown that the FDSS technique can be carried out on the time scales ranging from 2 to 160 ps and (a) yields 5-10 times better S/N ratio than pump-probe spectroscopy (PPS) for the same acquisition time, (b) suitable both for thick and thin samples, (c) can be used to study both the static and flowing samples, and (d) yields information on nonlinear refraction in addition to transient absorption. We describe a FDSS setup based on a variable-length grating compressor, in which the sandard instrumentation for spectral analysis is used (section IV). The proposed FDSS technique bypasses most of the problems associated with the spatial encoding and is easy to implement experimentally, as most of the components are already used for CPA (sections III and IV).

The paper is organized as follows: In section II and Appendix 1, we give a theoretical analysis of the FDSS technique. The requirements of the detection system and optics used for pulse chirping are scrutinized in section III. The details of a hybrid PPS-FDSS setup built in our laboratory are given in section IV. In section V and Appendix 2, FDSS kinetics for the selected photosystems are examined and compared to PPS kinetics obtained using the same hybrid setup. In section VI, the insights obtained in these experiments are summarized and the advantages and disadvantages of FDSS technique in our implementation are discussed. To save space, some figures (referred to in the text as, for example, Fig. 1S) and Appendices 1 and 2 are placed in Part II of this paper.

**II. THEORY**

Our treatment is similar in approach to the previous analyses of chirped pulse pump-probe TA spectroscopy [18,19] and interferometry;[17] however, it departs from these in more than one respect. Taking $\Delta\omega = \omega - \omega_0$ as the frequency offset for the probe pulse, the Fourier transform $E_\omega$ of the electric field $E(t)$ for a chirped Gaussian pulse centered at $\omega_0$ is given by

$$E_\omega = E_{probe} \exp\left[-\Delta\omega^2 \tau_p^2/2 - i\phi(\omega)\right] \tag{1}$$

where $E_{probe}$ probe is the time-integrated field and $\phi(\omega)$ is the frequency-dependent phase expanded as

$$\phi(\omega) \approx \frac{\phi''(\omega_0)}{2}\Delta\omega^2 + \frac{\phi'''(\omega_0)}{6}\Delta\omega^3 + ... \tag{2}$$

where the first term corresponds to the group velocity dispersion (GVD) and the second and following terms correspond to higher order dispersion. [9,18] The zero and first order terms in $\Delta\omega$ are omitted, because the linear term simply shifts the origin of time. Neglecting higher than quadratic terms and introducing a stretch factor $s = \phi''(\omega_0)/\tau_p^2$ and a complex "width" of the pulse $\tau_c^2 = \tau_p^2(1+is)$, we obtain $E_\omega \propto \exp(-\Delta\omega^2\tau_c^2/2)$. In the time domain, the oscillating field

$$E(t) = \frac{e^{-i\omega_0 t}}{\sqrt{2\pi}\tau_c} E_{probe} \exp(-t^2/2\tau_c^2) \tag{3}$$

has a Gaussian envelope with the pulse width of $\tau_p\sqrt{1+s^2} \approx \tau_p s$.

In the following, we will assume that the pump and the probe photons have vastly different energies and neglect coherent contributions to the TA signal, [21] as we are mainly interested in the dynamics that occur at delay times that are several times *longer* than the duration of the pump pulse. See ref. 18 for a more accurate derivation (for chirped pulse PPS) that takes into account both the sequential and Raman terms. [21] With our assumptions, the energy of the probe pulse absorbed in a thin sample of width $d$ is given by the Joule law

$$\langle U \rangle = \left\langle -E(t)\frac{d\Delta P(t)}{dt}\right\rangle \approx \frac{1}{2}\int_{-\infty}^{+\infty} d\omega\, \frac{\omega d}{c}\, \text{Im}\, E_\omega^*\, \Delta P_\omega \tag{4}$$

where $\Delta P(t) = \Delta\varepsilon(t)\otimes E(t)$ is the polarization induced by the pump pulse, $\Delta\varepsilon(t) = \Delta\varepsilon'(t) + i\,\Delta\varepsilon''(t)$ is the photoinduced change in the complex dielectric function of the sample and

$$\Delta P_\omega = \frac{1}{2\pi}\int_{-\infty}^{+\infty} dt\, e^{+i\omega t}\Delta\varepsilon(t)\otimes E(t) = \frac{1}{2\pi}\int_{-\infty}^{\infty} d\Omega\, E_\Omega \int_{-\infty}^{+\infty} dt\, e^{-i(\Omega-\omega)t}\Delta\varepsilon_\omega(t) \tag{5}$$

is the Fourier component of the third-order polarization. Since the change in the absolute frequency $\omega$ (corrected by the carrier frequency $\omega_0$) is small, the TA signal $S(\omega) = -\Delta T_\omega/T_\omega$, where $T_\omega$ is the transmission of the sample at the frequency $\omega$, is given by

$$S(\omega) = \frac{\langle \mathcal{C}\mathcal{S} \rangle_\omega}{\frac{1}{2}|E_\omega|^2} \tag{6}$$

Taking into account finite resolution of the spectrometer, eq. (6) may be rewritten as

$$S(\omega) = -2\ \mathrm{Re}\left[g(\omega) \otimes \frac{1}{t_\omega}\left(\frac{dt_\omega}{d\varepsilon}\right)\int_{-\infty}^{+\infty}d\Omega\ K_{\Omega-\omega}\ E_\Omega E_\omega^*\right] \Bigg/ \left[g(\omega) \otimes |E_\omega|^2\right] \tag{7}$$

where $t_\omega$ is the Fresnel transmission coefficient for the probe light of frequency $\omega$ (this generalization of eq. (4) is justified in Appendix 1),

$$g(\omega) = \frac{1}{\delta\sqrt{\pi}}e^{-\omega^2/\delta^2}, \tag{8}$$

is the resolution function of the detector (for the spectral power of $\delta$) which is convoluted with the numerator and the denominator of eq. (7) and

$$K_{\Omega-\omega} = \frac{1}{2\pi}\int_{-\infty}^{+\infty}dt\ \Delta\varepsilon_\omega(t)\ e^{-i(\Omega-\omega)t} \tag{9}$$

In the following, we will assume that the spectral profile of the photoinduced dielectric function $\Delta\varepsilon_\omega(t)$ is time independent. In this case, we can write $K_\mu = \Delta\varepsilon_\omega\ \tilde{K}_\mu$, where

$$\tilde{K}_\mu = \frac{1}{2\pi}\int_{-\infty}^{+\infty}dt\ \tilde{K}(t)\ e^{-i\mu t} = \frac{e^{i\mu T}P_\mu\ \Gamma_\mu}{P_{\mu=0}} \tag{10}$$

and $\mu=\Omega-\omega$. Function $\tilde{K}(t)$ in eq. (10) gives the formation and decay kinetics of the photoinduced species, where $P_\mu$ is the Fourier transform of the pump pulse (with the delay time $T$ given with respect to the center of the probe pulse) and $\Gamma_\mu$ is the Fourier component of the decay kinetics $\Gamma(t)$.

Eq. (7) was derived assuming that the transmission coefficient $t_\omega$ is a slow, nonoscillating function of $\omega$ in the spectral interval of $\omega_0 \pm 1/\tau_p$; otherwise (e.g., for thin film samples), more general eqs. (A18) and (A19) given in Appendix 1 must be used. For normal incidence of probe light on a flat thin sample, $t_\omega^{-1}\ dt_\omega/d\varepsilon = i\omega d/c$. For a very thin wedge (considered below), $t_\omega^{-1}\ dt_\omega/d\varepsilon = i\omega d/c\sqrt{\varepsilon_\omega}$. For a Gaussian pump pulse, $P(t) \propto |E_L(t)|^2 \propto \exp(-[t/\tau_L]^2)$, and exponential decay kinetics $\Gamma(t)=\exp(-\gamma t)$,

$$K_\mu = \Delta\varepsilon_\omega\ \frac{e^{i\mu T-\mu^2\tau_L^2/4}}{2\pi(\gamma+i\mu)} \tag{11}$$

For $g(\omega)=\delta(\omega)$ (infinite spectral resolution), eq. (11) simplifies to

$$S(\omega) \approx \frac{2\omega d}{c} \text{Im } \Delta n_\omega \frac{1}{2\pi} \int_{-\infty}^{+\infty} d\mu \frac{e^{i\mu T - \mu^2 \tau_L^2/4}}{\gamma + i\mu} \frac{E_{\omega+\mu}}{E_\omega} \tag{12}$$

where $\Delta n_\omega = \Delta \eta_\omega + i \Delta \kappa_\omega$ is the photoinduced change in the complex refraction index $n_\omega = \sqrt{\varepsilon_\omega}$ of the sample. If only the first term in eq. (2) is taken into account,

$$S(\omega) \approx \frac{2\omega d}{c} \text{Im } \Delta n_\omega \Phi(\alpha, \beta(\Delta\omega), \gamma) \tag{13}$$

where we introduced a function

$$\Phi(\alpha,\beta,\gamma) = \frac{1}{2\pi} \int_{-\infty}^{+\infty} d\mu \frac{e^{-\alpha^2 \mu^2 - \beta\mu}}{\gamma + i\mu} = \frac{1}{2} e^{\gamma^2 \alpha^2 - i\beta\gamma} \text{ erfc}\left(\gamma\alpha - \frac{i\beta}{2\alpha}\right) \tag{14}$$

with parameters $\alpha$ and $\beta$ given by equations

$$\alpha^2 = \frac{\tau_L^2}{4} + \frac{\tau_c^2}{2}\left(1 - \frac{\delta^2 \tau_c^2/2}{1 + \delta^2 \tau_p^2}\right) \tag{15}$$

$$i\beta = T + \frac{i \Delta\omega \tau_c^2}{1 + \delta^2 \tau_p^2} \tag{16}$$

Eqs. (14), (15) and (16) are generalized to include finite spectral resolution, i.e., non-zero $\delta$. For $\delta=0$, the expression is similar to eq. (18) [18] that was obtained for chirped-pulse PPS by Kovalenko et al. [18] In the following, we will assume that $\delta$ is much smaller than the width $1/\tau_p$ of the probe pulse in the frequency domain (i.e., $\delta\tau_p << 1$) and use the "delay time" $T_e$ defined as

$$T_e = -\text{Im } \beta = T - \Delta\omega\tau_p^2 s \tag{17}$$

For a sufficiently large frequency offset $\Delta\omega$ (long $T_e$), the error function in eq. (14) asymptotically approaches 2 and

$$S(\omega) \approx \frac{2\omega d}{c} \text{Im } \Delta n_\omega e^{\gamma^2 \alpha^2 - i\beta\gamma} \propto \frac{2\omega d \Delta\kappa_\omega}{c} e^{-\gamma T_e} \tag{18}$$

i.e., the signal $S(\omega)$ converges to exponential kinetics of photoinduced absorption, $\Delta\alpha_\omega(t) = \Delta\alpha_\omega \exp(-\gamma t)$, where $\Delta\alpha_\omega = 2\omega d\Delta\kappa_\omega/c$ is the absorption coefficient at the probe wavelength $\omega$ and $t=T_e$. Formula (17) can be generalized for higher dispersion orders by noticing that for a sufficiently small frequency $\mu$ (that gives the largest contribution to the integral in eq. (12)),

$$\frac{E_{\omega+\mu}}{E_\omega} = \exp\left(-\mu^2\tau_p^2/2 - \mu\Delta\omega\tau_p - i[\phi(\Omega) - \phi(\omega)]\right)$$
$$\approx \exp\left(-\frac{\mu^2}{2}\{\tau_p^2 + i\phi''(\omega)\} - \mu\{\Delta\omega\tau_p + i\phi'(\omega)\}\right) \quad (19)$$

Substituting this formula into eq. (12), the integral can be reduced to function (14) with

$$\alpha^2 = \frac{\tau_L^2}{4} + \frac{\tau_c^2}{2} + \frac{\phi''(\omega)}{2} \quad and \quad i\beta = T + i\Delta\omega\tau_p - \phi'(\omega) \quad (20)$$

so that the "delay time" $T_e$ is equal to the group delay at the frequency $\omega$ [18]

$$T_e \equiv -\operatorname{Im}\beta = T - \phi'(\omega) \approx T - \phi''(\omega_0)\Delta\omega - \frac{\phi'''(\omega_0)}{2}\Delta\omega^2 - \ldots \quad (21)$$

Since both parameters $\alpha$ and $\beta$ are complex functions of $\omega$, $S(\omega)$ strongly oscillates near the kinetics origin, where $T_e=0$ (the real and complex parts of function $\Phi$ are shown in Figs. 1 and 2). It was these oscillations that were observed by Beddard et al. [10] The oscillations are stronger for smaller $\tau_L$ and $\delta$, whereas longer pump pulses and lower spectral resolution damp these oscillations (see below). To characterize the oscillation pattern, we will assume that $T=\tau_L=0$ and $|s|\gg 1$. Then, for $\gamma=0$ (a step-like kinetics),

$$S(\omega) \propto \operatorname{Im} e^{i\phi_\varepsilon} \operatorname{erfc}\left(\frac{(1-i)}{2}\tau_{GVD}\Delta\omega\right) \quad (22)$$

where $\phi_\varepsilon$ is the phase of the complex refraction index $\Delta n_\omega$, and $\tau_{GVD} = \sqrt{|\phi''(\omega_0)|}$ ($= \tau_p\sqrt{|s|}$). Differentiating both sides of eq. (22) with respect to $\omega$ yields

$$\partial S(\omega)/\partial\omega \propto \sin(\Delta\omega^2\tau_p^2 s/2 - \pi/4 + \phi_\varepsilon) \quad (23)$$

Thus, the stationary points of $S(\omega)$ are symmetric about the origin at $T_e=0$ (Fig. 2). Introducing the time interval $\Delta T_\varepsilon^{(n)}$ between the $n$-th pair of these points in the group delay, we obtain

$$\Delta T_e^{(n)} = \tau_{GVD}\left\{2(\pi - 4\phi_\varepsilon + 4\pi n)\right\}^{1/2} \quad (24)$$

where the index $n=0,1,\ldots$ must be sufficiently large so that the square root exists. For the first pair of these stationary points,

$$\Delta T_e^{(1)} = \tau_{GVD}\sqrt{6\pi} \quad for \quad \phi_\varepsilon = \pi/2 \quad (25)$$

and

$$\Delta T_e^{(0)} = \tau_{GVD}\sqrt{2\pi} \quad for \quad \phi_\varepsilon = 0 \tag{26}$$

Thus, the oscillation patterns for photoinduced absorption ($\phi_\varepsilon=\pi/2$) and (nonlinear) refraction ($\phi_\varepsilon=0$) are quite different, both in their phase (eqs. (22)) and in the oscillation frequency (eq. (24) and Fig. 2). Since the error function becomes real for long "delay times" $T_e$, for $\phi_\varepsilon=0$ the nonlinear refraction contribution $S(\omega)$ tends to zero (the upper traces in Figs. 1 and 2). Thus, $S(\omega)$ is sensitive to nonlinear refraction only near the kinetics origin, where $T_e=0$. To see how the finite spectral resolution causes damping of $S(\omega)$, consider the case when the decay kinetics $\Gamma(t)$ is so fast that one lets $\gamma\to\infty$. In this case, we obtain

$$S(\omega) \approx Im\ \Delta\varepsilon_\omega \frac{1}{2\sqrt{\pi\gamma}\alpha}\exp\left(\frac{\beta^2}{4\alpha^2}\right) \tag{27}$$

For $T=0$ and $s\gg1$

$$\exp\left(\frac{\beta^2}{4\alpha^2}\right) \approx \exp\left(-T_e^2/\left\{\tau_L^2 + \delta^2\tau_{GVD}^4 - 2i\tau_{GVD}^2\right\}\right) \tag{28}$$

and $S(\omega)$ is damped as $\exp(-T_e^2/T_{damp}^2)$, where $T_{damp}$ is given by the real part of eq. (28). For $\tau_L\approx0$ and $\delta\tau_{GVD}<1$, $T_{damp}\approx2/\delta$. Thus, for $\delta=1$ cm$^{-1}$, the oscillations in the $T_e$ domain are damped in 10.6 ps (Fig. 4, traces (a) to (d) illustrate the oscillation damping for $\delta=2$, 5, and 10 cm$^{-1}$). For $\delta\approx0$, $T_{damp}^2 = \tau_L^2 + (2\tau_{GVD}^2/\tau_L)^2$, i.e., for $\tau_L > \tau_{GVD}/2$, at most one oscillation would survive (Fig 4, traces (e) to (h))

Using eqs. (13) to (16), one can find the optimum set of parameters needed to sample an exponential kinetics with a given time constant $\gamma^{-1}$:

$$|\Delta\omega_{max}| \approx 1/\tau_p,\ s \approx \pm2/\gamma\tau_p,\ and\ T \approx \pm1.25/\gamma \tag{29}$$

The first equation specifies the optimum spectral range for which $|E_\omega|^2$ is reasonably large at the extremes of this range. The next two equations specify the optimum stretch factor $s$ of the probe pulse and the pump delay time $T$ chosen to take the maximum advantage of the spectral range (where $s>0$ for a stretched pulse and $s<0$ for a compressed pulse). If only the kinetic profile of TA for $T_e \geq (2-5)\tau_{GVD}$ is of interest, the width $\tau_L$ of the pump pulse can be chosen so that $\tau_L \geq (0.5-1)\tau_{GVD}$; in this case the oscillations are almost completely eliminated (Fig. 4, traces (e) to (h)). To the same end, one can reduce the spectral resolution $\delta$ so that $\delta\tau_{GVD}>1$ (Fig. 4, traces (a) to (d)). If the oscillation pattern of $S(\omega)$ is of interest (e.g., for the measurement of $\Delta n_\omega$), one needs to minimize both $\tau_L$ and $\delta$.

In general, function $\Phi(\alpha,\beta,\gamma)$ must be evaluated numerically. This can be done by splicing a series expansion (for small $|z|$) and a continued fraction expansion (for large $|z|$) for the generalized error function $eerfc(z) = \exp(z^2)\,erfc(z)$ (eqs. 7.1.8 and 7.1.14 in

ref. 22). In our simulations, 200-500 terms were retained in both of the expansions. Alternatively, the integral in eq. (12) was evaluated numerically on a grid of $(1-5) \times 10^3$ frequency points. The latter method allows including higher dispersion orders, spectral truncation, etalon effects, etc.

Fig. 3 shows the real part of function $\Phi$ (eq. (14)) for several stretch factors $s$, pump pulse widths $\tau_L$, and spectral resolution parameters $\delta$ (Fig. 2). For picosecond kinetics ($s=100$), 30-50% of the time window is taken by the oscillation pattern; for larger GVD, this fraction decreases as $s^{-1/2}$. The observed trends agree with the qualitative analysis given above and the results in section V.2. Fig. 5 shows the progression of signals $S(\omega)$ as a function of the phase $\phi_e$ for $\Delta n_\omega = const(\omega)$. Observe how the oscillation pattern near the kinetic origin changes with the phase; this may be used to estimate $\phi_e$ from the general shape of the oscillatory pattern (section V.1). For $\Delta n_\omega$ that spectrally evolves in time this pattern reflects the average phase over a period of time $\sim \tau_{GVD}$. By changing GVD, one can obtain the time dependence of this average phase.

## III. EXPERIMENTAL CONSIDERATIONS.

### 1. Chirping the probe pulse.

As compared to CPA, FDSS sets very low constraints on the compressor or stretcher used to chirp the probe pulse. In the usual CPA scheme,[9] a short pulse from the oscillator is stretched, amplified, and then recompressed, so that the positive chirp added by the stretcher (and the amplifier optics) is cancelled by the negative chirp added by the compressor. Therefore, to obtain a transform limited, ultrashort pulse, this scheme requires compensation of third (TOD) and higher order dispersion. In the FDSS, TOD and the higher orders are unimportant because for the standard beam geometry, a grating compressor/stretcher cannot introduce sufficiently large TOD to cause notable distortion of the kinetics. For a single round-trip between the pair of parallel gratings, the GVD is given by[9]

$$\phi''(\omega_0) = -\frac{\lambda^3 L_g}{\pi c^2 d_g^2} \left[1 - \left(\frac{\lambda}{d_g} - \sin\theta\right)^2\right]^{-3/2} \quad (30)$$

and the TOD is given by

$$\xi_3 = \frac{\omega_0 \, \phi'''(\omega_0)}{3\phi''(\omega_0)} = -\left(\frac{\lambda}{d_g}\sin\theta + \cos^2\theta\right) \bigg/ \left[1 - \left(\frac{\lambda}{d_g} - \sin\theta\right)^2\right] \quad (31)$$

where $\lambda = 2\pi c/\omega_0$ is the wavelength at the center, $\theta$ is the angle of incidence on the grating (chosen to be as close as possible to the Littrow angle, $\sin\theta_L = \lambda/2d_g$), $L_g$ is the slanted distance between the gratings, and $d_g$ is the ratio of the groove spacing and the diffraction order $m$. For a stretcher, GVD has the opposite sign. Note that both GVD and $\xi_3$ are maximum at $\theta_L$, where $\xi_3 \approx -(1 + \lambda^2/4d_g^2)/(1 - \lambda^2/4d_g^2)$ is close to -1. We can rewrite eq. (21) as

$$T_e \approx T - \Delta\omega \ \phi''(\omega_0) \left[1 + \frac{3\xi_3}{2}(\Delta\omega/\omega_0)\right]. \qquad (32)$$

Since $\Delta\omega\tau_p<1$ in the optimum spectral range (eq. (28)), the correction to the second term of eq. (32) does not exceed $(1-2)\times(\omega_0\tau_p)^{-1}$ - which is at most a few per cent. Fig. 6a shows a simulation of $S(\omega)$ as a function of $T_e$ defined by eq. (17) for $\xi_3$=-1.6 (that corresponds to the Littrow angle dispersion of 800 nm probe light by a 1200 g/mm grating in the first diffraction order). While the TOD shifts the kinetic origin, the change in the kinetic profile is very slight. This small change can be completely eliminated by plotting the kinetics as a function of the group delay $T_e$ defined by eq. (32), as shown in Fig. 6b.

Another point is that the probe beam dispersed on the grating can be clipped. In CPA, the spectral wings cannot be clipped because the high-frequency components must be added to prevent oscillations of the compressed pulse in the time domain. By contrast, in FDSS, the kinetics are imprinted in the *central* section of the probe spectrum and Fourier components outside this section do not matter. To demonstrate this point, we used eq. (12) to calculate $S(\omega)$ for a probe pulse whose Gaussian spectrum was truncated at $\Delta\omega \approx \pm 1/\tau_p$ (Fig. 7). It is clear from this calculation that the effect of truncation is limited to narrow regions near the truncation points. This is explained as follows:

The clipping of the spectrum is equivalent to narrowing the integration interval over $\Omega$ (or $\mu$) in eq. (12). The integral function is damped as $\exp(-\mu^2\{\tau_L^2/4 + \tau_p^2/2\})$, and for $\tau_L$=100 fs pump and $\tau_p$=20 fs probe the Gaussian envelope is 160 cm$^{-1}$ FWHM. If the kinetic origin is removed from the truncation points by that much, the cutoff has no effect on the oscillation pattern. In general, the effect of limiting $\mu$ to a range of $(-M,+M)$ is equivalent to introducing an additional damping factor $\exp(-\mu^2/M^2)$ that increases $\alpha^2$ (given by eqs. (15) and (20)) by $M^{-2}$. For instance, a bandpass of $\pm 10$ cm$^{-1}$ is equivalent to increasing $\tau_L$ to 1 ps. Thus, a few picoseconds away from the kinetic origin, the truncation of the probe pulse has no effect since a very narrow band of the frequencies $\Omega$ contributes to $S(\omega)$ in the first place. Only for very low GVD (*s* of $\pm$(10-50)) can the truncation distort the kinetics. However, for those GVD's, the spread of the beam on the gratings is small and no clipping occurs anyway. This consideration indicates that once the kinetic origin is removed by 100-200 cm$^{-1}$ from the clipping points, the kinetics (including the oscillation pattern) do not change at all.

We conclude that for the same stretch factor *s*, the gratings in the FDSS setup can be placed 3-5 times further away than the gratings in the equivalent CPA design. A $\tau_p$=20 fs (33 fs FWHM), 800 nm pulse from a Ti:sapphire oscillator yields the optimum sampling interval (eq. (29)) of $\pm 250$ cm$^{-1}$. Simple estimates show that for the standard 1200 g/mm ($\theta_L$= 28.7°), 11 cm - wide reflection grating used in the first order, the distance $L_g$ can be increased to 1.5-2 m without adverse effects on the FDSS kinetics: For $\theta$=40° and $L_g$=1 m, $\phi''(\omega_0)$=-3 ps$^2$ (*s*=-7,650, $\xi_3$=1.34), and the $\pm 250$ cm$^{-1}$ interval would be dispersed over 4.5 cm. This calculation suggests that with the standard equipment, one can sample the kinetics over 300-500 ps.

It may occur that the use of smaller stretch factors (<$10^3$) and longer probe pulses (> 200 fs) might be preferable to compressing a 10-50 fs pulse by a large factor of -(1-10)x$10^3$, as done in our experiments (section IV). Since the oscillations take $\approx s^{-1/2}$ of the total kinetics regardless of the probe pulse width, larger GVD and shorter pulses actually yield "cleaner" TA traces, even for relatively long sampling times. Furthermore, resolving the oscillation pattern requires short pump pulses and, therefore, short probe pulses. In the end, the main (perhaps, the only) advantage of longer probe pulses is the possibility of stretching these pulses with optical fibers.

Since the spectral range and resolution are fixed, to obtain the kinetics on different time scales, GVD must be tuned – ideally, over the widest possible range (see, for example, section V.2). To this end, a variable-length grating compressor is best suited. In our design (section IV), the compression factor is changed by fixing the retroreflector RR2 (Fig. 8) and sliding grating GR2 using a 70 cm path translation stage (TS1). Though we used gold gratings, less expensive aluminum gratings can be used, as large power losses can be tolerated due to high intensity of the probe light. For the same reason, higher diffraction orders $m$ can be used to increase GVD without changing $L_g$ (eq. (30)). Since it is impossible to place large reflection gratings (needed to obtain large GVD) closer than 15-20 cm, an equivalent positive chirp was introduced prior to the compression. In our system, the probe pulse is taken from the amplified 800 nm light dispersed using an $L_g$=40 cm grating stretcher and then negatively chirped using a variable-length compressor (Fig. 8). A separate fixed-length compressor is used to generate a transform-limited, amplified 800 nm pulse that provides the pump light (Fig. 8). This scheme allows for continuous change in the probe GVD between 0 and -1.6 $ps^2$.

Alternative schemes for obtaining low stretch factors ($10^2$-$10^3$) include the use of transmission gratings (that can be put close to each other and, thereby, give the widest dynamic range),[23] dispersive materials (such as SF18 glass),[9] and optical fibers (especially for visible light).[10] For instance, SF18 glass has GVD of 1540 $fs^2$/cm at 800 nm ($s$=+3.85/cm, $\xi_3$=0.5)[9] and the stretch factor of +100 can be achieved by passing a transform limited probe pulse through 25 cm of this material. Beddard et al.[10] used a 2.5 m long single mode optical fiber to obtain the GVD of 0.14 $ps^2$ to perform their first FDSS experiment. Thus, one can envision a hybrid system where low (positive) stretch factors are obtained by using dispersive materials whereas large (negative) compression factors are obtained using compressors with reflection grating.

The greatest inconvenience presented by the FDSS technique is that the probe color cannot be easily tuned. The most popular method for varying the probe wavelength is by generation of white light supercontinuum and using narrow band interference filters to select a wavelength range. The supercontinuum is nonlinearly chirped, with a typical stretch factor of 20-100. Due to large beam divergence, it is difficult to compress these pulses using a gratings compressor (although, low GVD can be obtained by dispersion in a fiber).[10,18] Thus, one is forced to use optical parametric amplifiers (OPA's) as the source of coherent probe light. Furthermore, the compressor must be realigned for each probe color (due to the need to keep the incident angle reasonably close to the Littrow angle). While this operation can be automated, it is difficult to avoid large changes in the stretch factor since $s \propto \lambda^3$ (eq. (30)). A change-over from 800 nm to 400 nm decreases GVD by an order of magnitude. With the exception of low GVD's, this decrease cannot be compensated by the equivalent increase in $L_g$. Also, for $\lambda$ other than the oscillator wavelength there is no built-in source of stretched probe pulses, and the dynamic range is reduced.

It should be stressed that most of these problems present themselves for shorter wavelengths; for longer wavelengths, the change-over is less problematic. Even for shorter wavelengths, a ready solution exists provided that the probe beam is sufficiently bright to tolerate losses that occur when the gratings are used in the higher diffraction order. A typical OPA yields 2-10 µJ of the VIS light, whereas less than 1 nJ is needed to probe the sample. For a given groove spacing, the number of possible orders $m$ (given by $\lambda/d_g < 2$) increases as $1/\lambda$ whereas GVD increases as $m^2$, and the loss of GVD at the shorter wavelengths can be compensated by increasing $m$. E. g., for the standard 1200 g/mm grating compressor, 4 diffraction orders exist at 400 nm, and the Littrow angle for the third order is still acute, ca. 46°. With this grating used in the 2$^{nd}$ and 3$^d$ orders, the GVD at 400 nm is 0.5 and 2.3 times that at 800 nm in the first order. The chart shown in Fig. 3S helps to choose the optimum $L_g$ and $m$ to obtain a desirable stretch factor $s$ for a given wavelength $\lambda$, for a 33 fs FWHM pulse and 1200 g/mm grating (of the fixed width).

**2. The spectrum of the probe pulse.**

While the theoretical treatment of section II is formulated for an ideal Gaussian pulse, the spectra of probe pulses obtained using CPA (before or after the compression) often exhibit low-amplitude "ripple" (2-10%) juxtaposed on the Gaussian profile. The origin of this "ripple" can be traced to etalon and aperture effects during the passage of the beam through the amplification optics and to the modes of Ti:sapphire oscillator itself. While this "ripple" is reduced when the beam passes through a compressor, it reappears as the probe light is diffracted by the sample and the focussing optics. Due to this diffraction, the amplitude of the "ripple" is not uniform across the beam; the central part exhibits 2-3 times stronger oscillations than the outer part. A typical slit opening of the monochromator needed to obtain 1-5 cm$^{-1}$ resolution is 50-100 µm and, in the absence of a diffuser, it is the central part of the beam that is imaged on the detector. Since the reference and signal beams travel different paths and are diffracted differently, their "ripple" patterns are not identical. While most of the "ripple" is divided out by the normalization of the signal spectrum by the reference spectrum, better results are obtained when a diffuser (DF in Fig. 8) is inserted to homogenize the light. Though the "ripple" pattern is still observed, close semblance between the "ripple" patterns for the signal and reference beams is hieved.

While this "ripple" is a nuisance, it has almost no effect

o n　　　　　　　　　　t h e　　　　　　　　　　k i n e t i c s .

To the first approximation, the "ripple" can be simulated as

$$E_\omega = E_\omega^0 \left\{1 + A_{osc} \exp(i[\omega - \omega_0]/\omega_{osc})\right\} \qquad (33)$$

where $E_\omega^0$ is the $\omega$ component of the field without the "ripple", $A_{osc}$ (<<1) is the complex amplitude of the oscillations, and $\omega_{osc}$ is their period. For the field given by eq. (33), the ratio in eq. (12) is given by

$$E_{\omega+\mu}/E_\omega \approx E_{\omega+\mu}^0/E_\omega^0 \left\{1 + A_{osc} \exp(i\mu/\omega_{osc})\right\} \qquad (34)$$

and

$$S(\omega) \approx \frac{2\omega d}{c} \operatorname{Im} \Delta n_\omega \{\Phi(\alpha,\beta,\gamma) + A_{osc}\Phi(\alpha,\beta - i\omega_{osc}^{-1},\gamma)\} \approx$$
$$\frac{2\omega d}{c} \operatorname{Im} \Delta n_\omega [1 + A_{osc}] \Phi(\alpha,\beta,\gamma) \qquad (35)$$

The last approximation in eq. (35) is justified by the fact that offsetting the coefficient $\beta$ is equivalent to a shift of the delay time $T$ of the pump by $\omega_{osc}^{-1}$ - which is a small correction relative to the characteristic time $\tau_{GVD}$ of the oscillations shown in Figs. 1 to 5. We conclude that the high-frequency "ripple" is inconsequential; at most, it adds a small shift to the phase $\phi_\varepsilon$ of $\Delta\eta_\omega$.

## 3. Monochromator and detector.

Since a monochromator grating can always be chosen so that $|\Delta\omega_{max}| \approx 1/\tau_p$ is several times smaller than the spacing between the next diffraction orders, the use of low-density gratings in the higher order may be preferable to the use of high-resolution spectrometers with long focal distances and/or high-density gratings. Although the diffraction in higher order lowers the throughput, that rarely constitutes a problem unless white light supercontinuum is used (section III.2). Because the dispersion increases linearly with the diffraction order $m$, changing the order allows one to change the resolution power of the spectrometer without any other adjustments. Such flexibility is lacking for high-resolution spectrometers in which high-density gratings are used in the first order (as is done in most experiments). With our $f$=27 cm Czerny-Turner monochromator (SPEX model 270M) equipped with a 150 g/mm ruled grating (4 μm blaze), the spectral resolution $\delta$ of 2-5 cm$^{-1}$ (and dispersion of 3.1 nm/mm) in the eighth order was readily achievable for a 50-200 μm slit opening. For this order, the entire spectrum of a 33 fs FWHM pulse covers less than 20% of the angle between the orders. Equally good results were obtained using the same monochormator equipped with a 1200 g/mm grating (500 nm blaze) in the first order.

A typical multichannel analyzer has 500-1000 pixels over the flat field of the imaging spectrograph with a spacing of 20-25 μm per channel. In our setup, a thermoelectrically cooled 512 channel dual diode array (Princeton Instruments DIDA512 with 14 dynamic bits) was used. The signal on the diode array saturates at relatively high photon flux (> $10^8$ photons per pixel), and the digitization of the signal is more accurate as compared to a typical CCD (that saturates at < $10^6$ photons per pixel). On the other hand, the discharge of holding capacitors on a CCD takes < 1 ms vs. 40 ms on the intensified diode array, so the duty cycle is better on a CCD. These two factors balance each other at high repetition rates (1 kHz); for a low repetition rate (tens of Hz), the diode array has a clear advantage. For our 1 kHz Ti:sapphire system, typical duty cycle of the detector was 70%, with 60 ms exposure to the laser light per digitized frame. 5-10 such frames were averaged, and the series with the pump pulse off and on were alternated. A mechanical shutter (SH2 in Fig. 8) was used to block the pump light. The ratii of the probe and the reference signals were calculated for each acquisition channel, and the "pump on" ratii were normalized by the "pump off" ones to obtain $\Delta OD_\omega = -\log(T_{on}/T_{off}) \approx 2.3\ S(\omega)$ for each channel. 100-500 such "on/off" series were averaged to obtain the kinetic profiles shown in section V. Before the experiment, "pump on" and "pump off" dark signals (with the probe light blocked by another shutter (SH1 in Fig. 8) were collected and subtracted from the spectra. In a typical "blank" experiment with the pump beam blocked, the "on" and "off" series for the total of $3 \times 10^4$ pulses converged to a "$\Delta OD$" of $2 \times 10^{-5}$, the standard deviation across the spectrum of $5 \times 10^{-4}$; the "on" and "off" ratii were linearly correlated with $\rho^2 = 0.9996$. To estimate the flatness of the spectral response, a 1 mm thick 60% pass neutral density filter was periodically inserted into the "signal" beam using a motorized flipper mount (New Focus model 8892-K); the response was flat within ±3% across the spectrum. Similar estimates were obtained by insertion of a pellicle beam splitter and a 150 μm thick flat glass plate, from a fit to the observed interference pattern. It is difficult to improve on the flatness of the response due to the dispersion of probe light in the focussing optics and the sample.

### IV. EXPERIMENTAL.

The diagram of the setup is shown in Fig. 8. The pump and probe pulses were derived from a femtosecond CPA system: A diode-pumped Nd:YVO laser (Coherent Verdi, 5W) was used to pump Kerr lens mode-locked Ti:sapphire laser operating at 80 MHz (Spectra Physics Tsunami 3941). Single pulses were selected with a Pockels cell (Medox) and subsequently stretched to 80 ps in a 1200 g/mm grating stretcher. These 2 nJ pulses were then amplified to 0.7 mJ and then to 4 mJ in a two-stage multipass Ti:sapphire amplifier pumped with 8.5 mJ and 20 mJ light from diode-pumped Nd:YLF lasers (Spectra Physics Evolution 10 W and 30 W, respectively). The repetition rate of the amplifier was 1 kHz and the pulse to pulse stability was typically 3%.

The output from the amplifier was split 1:20 using a glass wedge. The main part of the beam was passed through a 1200 g/mm grating compressor that yielded Gaussian probe pulses of 50 fs FWHM and 3 mJ centered at 800 nm. These pulses were delayed using a retroreflector RR1 on a double path 99 cm motorized translation stage TS2 (Unislide model ZB40395J with a Velmex 86mm-2 controller) and doubled in frequency using a 400 μm Type I BBO ($\beta$-BaB$_2$O$_4$) crystal (SHG in Fig. 8). The typical maximum output of the 400 nm light was 80 μJ (100-200 fs FWHM). The 400 nm light was passed through a couple of narrow band 800 nm mirrors to remove the fundamental; in some instances, a small chirp on the pump pulse was introduced (to increase the product yield) by placing a 3 mm thick blue-green filter.

The probe pulse (800 nm, 10 mm diameter) was derived from the same amplifier output and chirped using a variable length compressor (1200 g/mm) with gold reflecting gratings. To this end, the larger (11 cm wide) grating GR2 was mounted on a 76 cm translation stage TS1 (Unislide model ZB253002J with a Compumotor S6 drive). An 18 cm wide, gold roof reflector (RR2) was fixed behind the smaller (6 cm wide) stationary grating (GR1). Note that this reflector should be sufficiently wide to compensate for the shift of the beam as GR2 is translated. The angle of incidence on GR1 for the 800 nm beam was 45°. In our design, the Fourier transform limited pulse was obtained for GR2 positioned 5 cm from the end of the stage, which gave us 55 cm of the slanted distance $L_g$ for compression. The spectrum of the compressed beam was continuously monitored using Ocean Optics model S2000 miniature fiber optics spectrometer (FBO) with a 1200 g/mm grating (0.3 nm FWHM resolution).

To characterize the chirp, the compressed probe beam was diverted with a dielectric mirror on a flip mount FM1 and used for autocorrelation using a 90 μm thin BBO crystal. A 20 mm path Melles Griot Nanomotion II stage was used to delay the 800 nm beam in one of the arms. A fast 1P28 photomultiplier (RCA) operated at -700 V was used to detect the 400 nm light from the BBO crystal; this signal was sampled using a 10 ns gated SR250 boxcar integrator. The photomultiplier was used since for large compression factors, the intensity of the doubled frequency pulses was too low for a photodiode. The pulse width of the compressed pulse obtained using this autocorrelator was in good agreement with the value estimated from the known $\tau_p$ (obtained from the spectrum width of the probe pulse) and GVD calculated using eq. (30). The calculated GVD was also in good agreement with the value obtained from eq. (17). For the latter calibration, the delay time of the pump pulse $T$ was incremented and the spectral shift of the kinetic origin for a photoinduced signal measured; the ratio of these two increments gives the GVD.

The main part of the compressed beam was passed through a 50% glass beam splitter (BS). One beam was used as a reference, another as a probe. A notch filter (NF) on a translation stage was placed before a 50% beam splitter to match the spectra of the two beams on the diode array (section III.3). Before the beam splitter BS, the 800 nm beam was passed through a 2-3 mm aperture (A) and attenuated down to < 1 nJ with a variable neutral density filter. To eliminate the possibility of coherent artifacts, the pump and probe pulses were perpendicularly polarized. The 800 and 400 nm beams were focused with achromatic lenses L1 and L2 ($f = 30\ cm$) and overlapped in the sample (SM) at 6.5°. The transmitted probe was collimated using a thin lens L3, and the "signal" and "reference" beams were used for either the standard pump-probe or for FDSS detection; dielectric mirrors on the flip mounts FM2 and FM3 were used to divert the beams to either one of the detection systems. In the pump-probe configuration, the signal and the reference signals were detected with fast silicon photodiodes PD1 and PD2 (EG&G FND-100Q biased at 90 V), amplified to 5 V, shaped to 1 μs width (using an EG&G 142A preamplifier and an EG&G 855 spectroscopy amplifier), and sampled with home-built sample-and-hold electronics. A Power Mac G3 computer was used to digitize the amplified diode signals using a 16 bit A/D converter (National Instruments PCI-MIO-16XE-50) and to control the diode array, shutters SH1 and SH2 (Uniblitz model

VS25S2ZMO), and the delay stages. A mechanical chopper (THz Technologies model C-980) locked at 50% repetition rate of the laser was used to block the pump pulses on alternative shots (this chopper was removed for FDSS measurements). The standard deviation for a single-pulse pump-probe measurement was typically $10^{-3}$, and the "noise" in the PPS kinetics was entirely dominated by the variation of the pump intensity and flow instabilities.

In the FDSS experiment, the probe and reference 800 nm beams were focussed on a diffuser (DF) that was placed 2-5 mm before the monochromator slit with a $f$=30 cm cylindrical lens (CL). See section III.3 for more detail on the acquisition methodology. Since nearly perfect wavelength correspondence between the signal and the reference signals is essential, the following alignment procedure was used: A notch filter NF (CVI model FI.5-794.7-0.5) centered at 794.2 nm with a 1.97 nm FWHM Lorentzian transmission profile was placed before the beams were split, and the spectra (dispersed over 150-200 photodiodes) fit with a Lorentzian. By moving the beams, the centroids and the widths of the two Lorenzians were matched to 0.4-1 pixels. The spectral resolution was determined in a separate experiment by observing the 546.1 nm mercury line.

*Materials.* A 3° angle, 2 mm thin wedge of uncoated polycrystalline ZnSe with surface flatness of ±0.25 μm was obtained from Janos Technology, Inc. (part A1505W129). A 1.4 μm thick film of undoped amorphous (a-) Si:H alloy (8 at. % of H) deposited on 1 mm suprasil substrate was obtained from Prof. H. Fritzsche of the University of Chicago (see ref. 24 for more detail). Nanopure water with resistance > 10 MΩ was used in the experiments with liquid samples; analytical grade NaI (Mallinckrodt) was used in the experiments on electron photodetachment from I$^-$. The experiments were carried out either with a 160 μm thick, 6 mm wide laser jet or 5 mm optical path suprasil flow cell. In the latter case, the pump and probe beams were crossed in the middle of the cell, to avoid burning the glass walls. In both cases, the linear velocity of the fluid was > 1 m/s. [25] All measurements were carried out at 295 K.

## V. PHOTOSYSTEMS.

### 1. Polycrystalline ZnSe.

ZnSe is a II-VI semiconductor with an optical gap of 2.67 eV. [26-29] This material strongly absorbs light with $\lambda$<550 nm but is transparent at 800 nm (60% transmission for the refraction index $\eta$ of ca. 2.5). [28] On the ns time scale, nonlinear refraction index and TA for ZnSe were studied using Z-scan and 4-wave mixing spectroscopies that utilized 532 nm and 1064 nm probe pulses. [28] More recently, ultrafast time-resolved Z-scan spectroscopy has been carried out in the pump-probe fashion. [26,27] These data suggest that on the ps-to-ns time scales, $\Delta\eta$ is *negative* and dominated by free-carrier refraction; the *positive* signal due to the thermal effect is small. [27,28] Within the duration of a below-the-gap 532 nm pump pulse, $\Delta\eta$ is dominated by a short-lived *positive* bound-electron contribution; this contribution is lacking for above-the-gap excitation with a 395 nm pump. [27] For the latter, $\Delta\eta<0$ for all delay times and the refraction signal from the free carriers decays biexponentially, with the time constants of 7 and 63 ps. [27] The slow component is due to the charge recombination; the fast component is due to the trapping of the free carriers by defects and impurity. The same two components were observed using time-resolved luminescence spectroscopy. [27] The thermalization of free carriers is much faster than either one of these two processes. [27,28] For *n*-type ZnSe film excited with 420-480 nm photons, relaxation times of 0.5 and 1.8 ps were obtained for electron relaxation by emission of LO phonons and intervalley scattering, respectively. [29] A differential reflectance measurement for 400 nm photoexcitation also gave 0.44 ps for the electron relaxation. [30] Thus, we may safely assume that the oscillation pattern of $S(\omega)$ for $T_e$<10 ps is dominated by negative nonlinear refraction from thermalized photocarriers. For these carriers, we may use a simple Drude model [31,32] with

$$-\frac{\Delta\eta_\omega}{\eta_\omega} = \frac{\omega\tau_D \Delta\kappa_\omega}{\eta_\omega} = \frac{\omega_p^2/2}{\omega^2 + \tau_D^{-2}} \qquad (36)$$

where $\eta_\omega$ is the refraction index of the material, $\omega_p = (Ne^2/\varepsilon_0 m^*)^{1/2}$ is the plasma frequency of the carriers, $N$ is their number density, $\tau_D$ is the mean scattering time (assumed to be equal for the electrons and holes), $e$ is the elementary charge, and $m^*$=0.12 $m_e$ is the reduced effective mass of the carriers. [28] From eq. (36), we immediately obtain that $\cot\phi_\varepsilon \approx -\omega\tau_D$. Since the oscillation pattern of $S(\omega)$ uniquely depends on the phase $\phi_\varepsilon$ (section II), FDSS measurement yields $\tau_D$ without separate measurements of the absorption and refraction in the photoexcited sample. Note that common pump-probe approaches to the measurement of $\phi_\varepsilon$ would be (i) the combination of transmission and reflection measurements, [31,1S] (ii) the combination of two reflection measurements at different angles of incidence, [32] and (iii) time-resolved interferometry. [17]

Fig. 4S(a) exhibits the pump-probe kinetics obtained upon 400 nm excitation of a polycrystalline ZnSe sample. The pump beam was < 1 μJ to prevent the damage to the sample, so $\Delta OD$ at 800 nm was low, ca. 2.5x10$^{-2}$. A biexponential fit with time constants of 13.7 and 913 ps was used to fit these kinetics. The slower component becomes faster at the higher pump radiances which suggests that this component is due to the recombination dynamics, whereas the faster component is from the carrier trapping. [27] Fig. 4S(b) exhibits FDSS kinetics from the same sample that were obtained for $\tau_p$=20 fs, $\tau_L$=100 fs, GVD of -0.82 ps$^2$ ($s$=-2,048) and $\tau_{GVD}$=905 fs (75K pulses were averaged to achieve a standard deviation of 3x10$^{-4}$ on the base line). In Fig. 9(a), these FDSS kinetics were normalized by the $T$=+30 ps "kinetics" obtained to determine the spectral response of the sample to the probe light (that was not quite flat, see Fig. 4S(b)). Most of the signal is comprised of the "spike" near the origin that exhibits several well-resolved beats; the plateau is 4 times weaker than the central beat. In Fig. 9(a) this plateau is compared in shape with the kinetic profile obtained in the pump-probe experiment. These two kinetics are identical within the experimental error. Note that the FDSS signal for $T_e >> \tau_{GVD}$ is 5 times lower than the PPS signal, indicating that the pump-probe measurement was actually that of nonlinear refraction (in agreement with the results of refs. 27, 28, and 29). The same is suggested by the fact that the spacing between the largest (zero-order) oscillation in Fig. 9(b) given in the units of $\tau_{GVD}$, is 2.7 which is closer to 2.51 given by eq. (26) than to 4.34 given by eq. (25). Fig. 9(b) shows the least squares fit of the kinetics using eq. (12) (modified to include the two exponential components) in which the phase $\phi_\varepsilon$, the weights of the two components, and the time constant of the faster component were floating parameters. This fit gave a time constant of 11±1 ps for the fast component (which is reasonably close to 13.7 ps obtained from the pump-probe experiment) and $\phi_\varepsilon$=171±0.5°. From this phase, we estimate $\tau_D \approx 2.8$ fs, which is in the correct range for low-density electron plasma in a polycrystalline semiconductor. [31,32]

This example illustrates two points. First, using FDSS method one can, at a glance, determine whether the signal is dominated by TA or nonlinear refraction. Second, from the oscillation pattern, one can estimate the phase of the dielectric function. Simultaneously, TA kinetics can be determined from the "tail" of the FDSS kinetics. Although the S/N ratio for the FDSS kinetics shown in Fig. 9 was inferior to that for the PPS trace in Fig. 4S(a) obtained under identical excitation conditions, the acquisition time was 3 times shorter for the former. Also, in this particular pump-probe "absorption" measurement the PPS signal was 5 times (!) larger than the actual TA because it was mainly from the photoinduced change in the refraction index.

## 2. Biphotonic charge-transfer-to-solvent (CTTS) in aqueous sodium iodide.

Electron photodetachment from aqueous halide anions is a popular system for modeling solvation-driven electron transfer and, as such, it has been extensively studied in recent years. [33-37] The CTTS band of $(I^-)_{aq}$ is centered at 5.5 eV above the ground state, and the electron can be detached from this anion using one- or two-photon excitation with 225-272 nm [34,35,36] or 310-313 nm [33] photons. The photoreaction results in the formation of a neutral iodine atom, $(I^0)_{aq}$, that does not absorb in the VIS and NIR and the so-called "hydrated" electron, $e^-_{aq}$, that has a broad absorption band across the 300-1300 nm range which is centered at 720 nm (at 25°C). [34] This band corresponds to the bound-to-bound (1s→2p) transition of the electron localized in a nearly spherical potential well. Since the absorption band of the hydrated electron is much broader than the spectral width of the 800 nm probe pulse, this spectrum is "flat" over the sampling range and shows little time evolution after the first 5 ps (see below).

According to the model of Kloepfer et al. [35] (which is, in turn, based on the molecular dynamics simulations of Staib and Borgis and Rossky and co-workers, see refs. 36 and 37), the electron generated by CTTS is preferentially injected into a shallow (77 meV) trap that resides 0.5 nm away from the parent anion, forming a metastable $(I^0:e^-_{aq})$ pair; [35] the charge separation is over in < 200 fs. This pair is stabilized by weak attraction between the electron and the readily polarizable iodine atom. Between 0.2 and 2 ps, the solvent relaxes around the electron causing a continuous blue shift of the TA band. [34,36] At later delay times, the electron either recombines with the iodine atom with a time constant of ca. 33 ps (at 295 K), or escapes into the bulk with a time constant of ca. 70 ps. [35]

Figs. 5S and 6S show pump-probe TA kinetics for $e^-_{aq}$ probed at 800 nm following 400 nm excitation of 75 mM iodide in water. Trace (i) in Fig. 6S(a) was obtained with a transform limited 800 nm probe pulse; trace (ii) was obtained using the same pulse compressed to 40 ps. For pump radiance < 0.3 TW/cm$^2$, this excitation is biphotonic (Fig. 5S) [37] (our estimate for the absorption coefficient is 9.65±0.1 cm/TW) and yields sub-ns kinetics (shown in Fig. 6S(a)) that are similar to the kinetics reported by Kloepfer et al. [35] for bi- 255 nm photon excitation, suggesting similar photophysics. These kinetics can be (formally) fit using a biexponential dependence with time constants of 18.3 and 362 ps. For pump irradiance > 0.3 TW/cm$^2$, the kinetics become considerably flatter as the pump power increases (Fig. 5S(b), traces (i) and (ii)) and the photoinduced signal from $e^-_{aq}$ increases *linearly* with the irradiance (this dependence is shown in Fig. 5S(a)). The same behavior, in the same power range, was observed for neat water (see the next section and Fig. 10S(a)); we attribute this switch-over to 400 nm excitation of photogenerated pre-thermalized electrons into the conduction band of water occurring within the duration of the excitation pulse. This photoexcitation increases the initial separation between the thermalized electron and the iodine atom and slows down their geminate recombination which can be observed on the sub-ps time scale. Since optical cells are rapidly destroyed by the terawatt radiation, at these high pump irradiances the TA experiment have to be carried out in a high-speed jet whose wobbling surface adds to the signal instability.

The PPS traces shown in Fig. 6S(a) were obtained over 30 min period for a sample flowing in a glass cell (2-photon excitation). Still, due to the variations in the pump intensity, the S/N ratio was 20:1. Furthermore, nearly 2/3 of the scan time was used to translate the delay stage. The "noise" is even worse in the "2+1" regime (Fig. 5S(b)), in which the amplitude is more sensitive to the variation of the 400 nm pump intensity.

For FDSS, the chirp was optimized for the fast component, so that GVD was ca. -0.8 ps$^2$. Fig. 6S(b) shows FDSS kinetics obtained for several delay times $T$ of the pump (30K shots per trace). The peak $\Delta OD$ signals (ca. 0.14) obtained using the PPS and FDSS are almost the same, suggesting that most of the signal is due to *photoinduced absorbance* of the solvated electron. These kinetics were normalized by the $T$=240 ps trace (for which the transmission spectrum is least affected by the slow decay), and the kinetics shifted in time accordingly to obtain a spliced trace shown in Fig. 7S(a) (the PPS kinetics are juxtaposed on these FDSS kinetics). It is seen that the FDSS and PPS traces coincide within the 95% confidence level of the PPS trace, while the S/N ratio of the FDSS kinetics is much better, despite the fact that it took us as much time to obtain all six of these kinetics as to obtain the single pump-probe kinetics. Actually, the "noise" is a lesser problem than nonflat spectral response, part of which is due to the slow component in the relaxation dynamics of hydrated electron that was not observed by PPS, due to insufficient spectral resolution. Fig. 8S shows normalized FDSS kinetics (obtained for the maximum GVD of -1.58 ps$^2$) at different irradiances; each of these kinetics was obtained over a 5 min sampling period. The improvement in the S/N ratio, as compared to the PPS kinetics shown in Fig. 5S(b), is remarkable, given that both sets of the kinetics were obtained in a thin 150 μm jet, just below the dielectric breakdown. The change in the kinetic shape with the pump power reflects a switchover between the 2- and "2+1" photon excitation (see Fig. 5S(a)).

In the FDSS kinetics shown in Fig. 6S(b) and 7S(b) there is a "spike" on the rising edge of the $\Delta OD$ signal, where $T_e<T$ (indicated with arrows), that does not have a symmetric counterpart for $T_e>T$. Fig. 10 (to be compared with the theoretical curves shown in Fig. 3) demonstrates the evolution of the oscillation pattern as a function of a stretch factor; in Fig. 9S the same patterns are replotted vs. the reduced time, $(T_e - T)/\tau_{GVD}$, to facilitate the comparison. It is apparent from Figs. 10 and 9S that the oscillation pattern is asymmetric about the time origin; this asymmetry progressively decreases as $\tau_{GVD}$ increases from 0.15 to 1.2 ps (Fig. 9S), due to the relative decrease in the "spike" amplitude. For the lowest stretch factor of -60, the "spike" is nearly as large as the signal itself.

A "spike" with a symmetric counterpart near the top of the signal could have been explained by positive nonlinear refraction in the sample (see Fig. 5). This, however, cannot be the case since at longer delay times, FDSS signal almost equals the absorbance (as determined by the PPS). On the other hand, the "spike" oscillation takes 1-3 ps, depending on the GVD, and nonlinear refraction that contributes to the oscillation pattern cannot be explained by impulsive Raman scattering that occurs within the duration of a 200 fs FWHM pump pulse. The most likely origin of this feature is the spectral evolution that occurs on time scales less or comparable to $\tau_{GVD}$. This rapid evolution has been observed in the PPS experiments [33-37] and modelled [36] as a continuous red shift of the electron band as the species thermalizes. The theory developed in section II does not deal with such a situation; nevertheless, some inference can be made as to the effect of such an evolution. In particular, if the thermalization process takes less than $\tau_{GVD}$, the distortion of the oscillation pattern will be limited to at most one oscillation with a period of $\approx \tau_{GVD}$, and this oscillation will precede the pattern from the thermalized species in the group delay. This is precisely what is observed in Figs. 10 and 9S. For a Lorentzian line with a center at $\omega=\omega_c$ and half-width of $\Lambda$, $\cot\phi_\varepsilon \approx (\omega_c - \omega)/\Lambda$. As the center $\omega_c$ of the $e^-_{aq}$ band (with a nearly constant $\Lambda\approx 3500$ cm$^{-1}$) shifts from 10,000 cm$^{-1}$ to 13,890 cm$^{-1}$ during the solvation process, [36] the phase $\phi_\varepsilon$ swings from $-(50-70)^o$ to $+68^o$ (at which $\sin\phi_\varepsilon \approx 0.95$). It is this rapid phase change that is "imprinted" in the asymmetric oscillation pattern. As shown in the next section, a similar feature is observed for hydrated electron generated by photoionization of *neat* water.

To conclude, the iodide system illustrates the superiority of FDSS over PPS for a photosystem where the "noise" is mainly due to poor stability in the yield of photogenerated species. Not only were we able to correctly reproduce the kinetics in the selected time window; it was possible to splice several such kinetics to cover a longer delay range. Evidence for the predominance of absorption in the $\Delta T_\omega/T_\omega$ signal was obtained and short-term spectral evolution recognized through the asymmetry of the oscillation pattern.

### 3. Photoionization of neat water.

The conduction band of liquid water is positioned 8-9 eV above the ground state and simultaneous absorption of three 400 nm photons is sufficient to ionize this liquid. [38] A 3-photon absorption coefficient of 900 cm$^3$/TW$^2$ has been reported for H$_2$O; [39] recently, we revised this estimate to 270 cm$^3$/TW$^2$ and obtained a quantum yield of 0.41 e$^-_{aq}$ per 3 photons (unpublished). When the radiance exceeds 0.6-1.2 TW/cm$^2$, another 400 nm photon is absorbed by a pre-thermalized electron and the resulting "hot" electron is injected deep into the conduction band (Fig. 10S). In less than 50 fs, this electron localizes several nanometers away from the parent hole. As a result, the "3+1" process makes the geminate recombination less efficient. The escape yield of hydrated electron at 500 ns increases from 70% to 92% (see, for example, Fig. 10S(b)) as the average width of the electron-hole distribution increases from 1.15 nm (for the 3-photon process) to 2.7 nm (for the "3+1" process). [38]

We chose this photosystem because it provides the worst scenario both for pump-probe and spatial-coding "single-shot" spectroscopies: The high, mixed photon order of the ionization causes strong shot-to-shot variation in the electron yield; this yield also strongly varies across the excitation beam. For the average pump radiance of 0.1-0.4 TW/cm$^2$, a switchover of the dynamic behavior from 3 photon to "3+1" process occurs at the higher end of the variation range. Since water and most other dielectric materials break down at 1-5 TW/cm$^2$, the photolysis is carried out in a high-speed jet, and the light scatter and thermal lensing in the jet further add to the noise. The absorption of the 400 nm light in water is extremely non-uniform; in particular, at 0.5-1 TW/cm$^2$, 80-90% of the light is absorbed in a 10-20 μm thin layer near the surface. Although $\Delta OD$ of 0.01-1 can be obtained for radiances just below the dielectric breakdown (10 TW/cm$^2$), the shot-to-shot variance of the TA is comparable to the TA signal itself.

Pump-probe kinetics of hydrated electron in neat H$_2$O were obtained in the 3-photon regime using 0.15 TW/cm$^2$, 400 nm pump pulses of 200 fs FWHM. The pump radiance was estimated from the maximum $\Delta OD$ and the known photophysics parameters. Fig. 11S(a) shows FDSS kinetics normalized by the spectral response determined at later delay times (in the same way as for the iodide system). For $s$=-2,130, the maximum time window is ca. 60 ps, and the decay kinetics were quite flat. The oscillation pattern near the kinetic origin exhibits an asymmetric pattern that is similar to the one found for $e_{aq}^-$ in the iodide system; perhaps, it also originates in the solvation dynamics of the electron on the fast time scale. Fig. 11S(b) shows 140 ps long stretches of the FDSS kinetics obtained at GVD of -1.46 ps$^2$ for three pump powers in the "3+1" region. As $\Delta OD$ increases from 0.017 to 0.2 to 0.4; the amplitude of the oscillation pattern decreases in the opposite order. Similar trends were observed for other photosystems (e.g., see Fig. 8S): the interference effects that give rise to the oscillation pattern were much weaker in these strongly absorbing samples.

Note that same "noise" level was observed in the FDSS kinetics before and after the pump pulse, suggesting that most of the "noise" was due to the flow instabilities in the jet, despite the high photon order. Though the S/N ratio for the FDSS kinetics obtained using the jet was significantly worse than the same ratio obtained for solid samples and liquid solutions flowed in a cell, the averaging was still considerably more efficient than in the PPS experiment. We believe that most of the scatter is caused by a wobbling jet surface: Near the nozzle, where the flow is most stable, the jet is a concave lens that strongly refracts and disperses probe light.

## VI. CONCLUDING REMARKS.

In section V and Appendix 2, several photosystems were studied using FDSS in which the probe light was chirped using a grating compressor. In all four of the examined photosystems, FDSS yielded a superior S/N ratio than PPS over a shorter acquisition time. The sampling time was reduced due to the "single-shot" nature of the method and elimination of (a relatively slow) mechanical movement of the translation stage. Signals as low as 10$^{-3}$ can be studied using this "single-shot" technique.

FDSS showed another advantage, as it allowed us to recognize whether the photoinduced change in the transmission of the sample was due to TA or nonlinear refraction. For free carrier signal in ZnSe, the phase of $\Delta\varepsilon_\omega$ of the complex dielectric function was determined with an accuracy of a few degrees, directly yielding the scattering time of free electron plasma in this material (section V.1). In Appendix 2, we illustrated the use of FDSS for a thin sample that exhibited well resolved interference fringes whose spacing was similar to the spectral width of the probe pulse – an especially vexing case for PPS. In sections V.3 and V.4, TA kinetics for 2-to-4 photon ionization of a liquid sample (in a flow cell and in a high-speed jet) were obtained. The S/N ratio for these kinetics greatly exceeded this ratio for PPS kinetics collected over 5-10 times longer acquisition time. As shown in section III.1, the time window (< 160 ps for our setup) of FDSS can be extended to 300-500 ps, with 500-1000 channels across the sampling interval. These FDSS kinetics can be spliced together to obtain even longer time profiles (section V.3). On the other hand, kinetics shorter than a few picoseconds exhibit strong oscillations (Fig. 10) and may be of limited use without analysis.

We conclude that FDSS technique is widely applicable, versatile, and easy to implement experimentally. The frequency domain spectrometer can be built side-by-side with a pump-probe spectrometer in a single setup (section IV). The requirements for the compressor are not stringent and can be met with the standard equipment used for chirped-pulse amplification. Long-arm double pass compressors with inexpensive gratings in higher diffraction order can be used to cover a wide spectral range without much loss in the GVD (section III.1). The requirements on the detector system are minimal and can be met without recourse to high-resolution spectrographs (section III.3). Finally, the method does not require tight control over the profile of the probe and pump beam (and *a priori* profiling of the latter), as is the case with "single-shot" methods based on spatial encoding. [2-6] In particular, the detection can be performed far from the sample, which makes it ideal for use in confined spaces. As for the spectral profile of the probe pulse, it should be reasonably close to a Gaussian; however, a "ripple" pattern with peak-to-peak amplitude of 10-20% can still be tolerated (section III.2). For applications where amplitude variations of the kinetics are large and repetition rates are low, FDSS has clear advantage over PPS and most other "single-shot" spectroscopies. Though we have discussed transient absorption only, the implementation of FDSS for transient reflectance is straightforward.

These many advantages come with a grain of salt: FDSS kinetics are not easy to interpret when the photoinduced spectrum is changing with time. PPS yields kinetics that are averaged over the entire band pass of the probe pulse; even if there is a spectral evolution within this bandpass, the acquired dynamics can still be used to make an inference about the system. By contrast, to recover $\Delta\varepsilon_\omega(t)$ from FDSS kinetics one needs to know *exactly* how the optical spectrum evolves in the given bandpass. This confines the technique to photosystems that exhibit relatively broad TA spectra (that are featureless over 200-500 $cm^{-1}$). Narrow band signals (20-100 $cm^{-1}$) can still be studied using FDSS provided that the probe pulse is relatively long (in the picosecond range). With that deficiency in mind, we are confident that FDSS will find many potential uses for ultrafast research.

## VII. ACKNOWLEDGMENT

We thank Prof S. Bradforth, Drs. D. Gozstola, D. M. Bartels, and C. D. Jonah for many helpful discussions. This work was performed under the auspices of the Office of Basic Energy Sciences, Division of Chemical Science, US-DOE under contract No. W-31-109-ENG-38. SP acknowledges the support of the DGA through the contract number

DSP/01-60-056.

**Figure captions.**

Fig. 1

Real ($\Phi'$) and imaginary ($\Phi''$) part of function $\Phi(\alpha,\beta,\gamma)$ introduced in eq. (14) plotted as a function of the frequency offset $\Delta\omega = \omega - \omega_0$ (to the bottom) and the group delay $T_e$ (to the top; eq. (17)). The real part corresponds to the contribution from transient absorption; the imaginary part corresponds to the contribution from nonlinear refraction. The dash-dot bell-like trace is the spectrum of the probe pulse, $|E_\omega|^2$. The dotted trace is the exponential kinetics convoluted with the Gaussian pump and probe pulses. The following parameters were assumed:
$\tau_p = 20\,fs$, $\tau_L = 100\,fs$, $s = 2048$, $\Delta = 2\,cm^{-1}$, $T = 30\,ps$, and $\gamma^{-1} = 40\,ps$.

Fig. 2

A closer look at the oscillation pattern for the real and complex parts of function $\Phi$ (eq. (14) for $\tau_p = 20\,fs$, $s = 2048$, and $T = \gamma = 0$. Solid traces are for $\tau_L = 100$ fs and $\delta = 2$ cm$^{-1}$; dotted traces are for $\tau_L = \delta = 0$. The group delay is given in the units of $\tau_{GVD} = \tau_p \sqrt{s}$ (equal to 905 fs). The extrema for $\Phi'$ correspond to the saddle points for $\Phi''$, and *vice versa*. The positions of the extrema are given by eqs. (24), (25), and (26).

Fig. 3

Real part of function $\Phi$ (eq. (14)) as a function of the stretch factor: (a) $\gamma^{-1} = 1$ps, $T = 1.25$ ps, and $s = 100$ (GVD of $4 \times 10^4$ fs$^2$), (b) $\gamma^{-1} = 5$ ps, $T = 6.25$ ps, and $s = 500$ (GVD of $2 \times 10^5$ fs$^2$), and (c) $\gamma^{-1} = 50$ ps, $T = 62.5$ fs and $s = 5000$ (GVD of $2 \times 10^6$ fs$^2$). Other calculation parameters: $\tau_p = 20$ fs, $t_L = 100$ fs, and $\delta = 2$ cm$^{-1}$. The higher the GVD, the smaller section of the kinetics exhibits the oscillations.

Fig. 4

Real ($\Phi'$) part of function $\Phi(\alpha,\beta,\gamma)$ introduced in eq. (14) plotted as a function of the frequency offset $\Delta\omega$ (bottom) and group delay $T_e$. Save for the spectral resolution $\delta$ and pump pulse duration $\tau_L$, the parameters are the same as in Fig. 1. For traces (a-d) $\tau_L = 0$ and $\delta = 0$ (a), 2 cm$^{-1}$ (b), 5 cm$^{-1}$ (c) and 10 cm$^{-1}$ (d), respectively. For traces (e-h) $\delta = 0$ and $\tau_L = 100$ fs (e), 300 fs (f), 500 fs (g) and 1 ps (h), respectively.

Fig. 5

Simulated $S(\omega)$ kinetics (plotted against the group delay $T_e$) as a function of phase $\phi_\varepsilon$ of a frequency-independent photoinduced change $\Delta n_\omega$ in the complex refraction index (same parameters as in Fig. 1). The phase in degrees is indicated next to the traces.

Fig. 6

Effect of nonzero third order dispersion (TOD) on the frequency-domain kinetics $S(\omega)$ for pure photoabsorption ($\phi_\varepsilon=90^\circ$). Same parameters as in Fig. 1 except for $\delta=0$. The dashed trace for $\xi_3=0$ was obtained analytically using eq. (14); the solid trace for $\xi_3/\omega_0\tau_p = -0.034$ (calculated using eq. (31) for Littrow angle dispersion of 800 nm pulse on a 1200 g/mm grating in the first order) was obtained by numerical integration of eq. (12). In (a), both traces are plotted as a function of the group delay $T_e$ calculated for GVD only (eq. (17)). In (b) the trace for nonzero TOD is plotted vs. the group delay time $T_e$ given by eq. (32).

Fig. 7

Effect of truncation of the probe spectrum on the $S(\omega)$ kinetics (same parameters as Fig. 1 except for $\delta=0$). The dashed curve shows the clipped wings of the probe pulse spectrum. The bold line indicates the $\pm 250$ cm$^{-1}$ section. The thin solid line is the kinetics obtained with the truncated pulse (by numerical integration of eq. (12)); the dotted line is the same kinetics obtained with the full spectrum. The two kinetic traces are the same except for the two narrow regions at the truncation points.

Fig. 8

The optical scheme of the setup. See sections 3 and 4 for more detail.

Fig. 9

(a) Traces (i) to (v) are FDSS kinetics obtained for a 1 mm thick polycrystalline sample of ZnSe excited with 400 nm photons and probed with a compressed $\tau_p$=20 fs, 800 nm probe pulse ($s$=-2,048, $\tau_{GVD}$=905 fs). See also Figs. 4S(a) and 4S(b). These FDSS traces were obtained for several delay times $T$ (0, -10, -20, -30, and -50 ps, respectively) of the $\tau_L$=100 fs pump pulse and normalized by the $T$=+30 ps trace (not shown) that yielded the spectral response of the sample (Fig. 4S(b)). Filled circles correspond to the PPS kinetics shown in Fig. 4S(a) that were scaled down by a factor of 5 to match the FDSS kinetics, trace (i), in amplitude. In the plateau region of the FDSS traces, $\Delta OD$ signal is ca. $3 \times 10^{-3}$. The large scaling factor and poor S/N ratio in the FDSS traces originate through the fact that most of the "transient absorbance" signal obtained using PPS was due to the photoinduced negative change in the refraction index. (b) A simlation of the oscillation pattern near the kinetic origin (trace (i)) using the equations derived in section II. The simulation parameters are given in the text.

Fig. 10
Traces (a) to (d): 512 channel FDSS kinetics obtained upon biphotonic 400 nm excitation of 75 mM aqueous iodide flowing in a 5 mm optical path silica cell. The delay time $T$ of the pump pulse was chosen to take the maximum advantage of the 800 nm probe bandpass and differed between the traces. The 800 nm absorbance is from hydrated electron, $e_{aq}^-$, that fully thermalizes in 2 ps. The stretch factor $s$ and $\tau_{GVD}$ (for a $\tau_p$=20 fs probe pulse) are, respectively (a) -59 and 153 fs, (b) -370 and 384 fs, (c) -1,430 and 756 fs, and (d) -3,720 and 1.22 ps. Each trace is the average of 90K pump on - pump off shots. The maximum FDSS signal is ca. 0.1; the spectral resolution $\delta$ is 2 cm$^{-1}$. The arrows indicate the position of the "spike" that lacks a symmetric counterpart for $T_e > T$ (see the text). See Fig. 9S for these oscillation patterns replotted as a function of the reduced time, $(T_e - T)/\tau_{GVD}$, to facilitate the comparison of the features observed for different compression factors. Compare these experimental traces with the theoretical traces shown in Fig. 3.

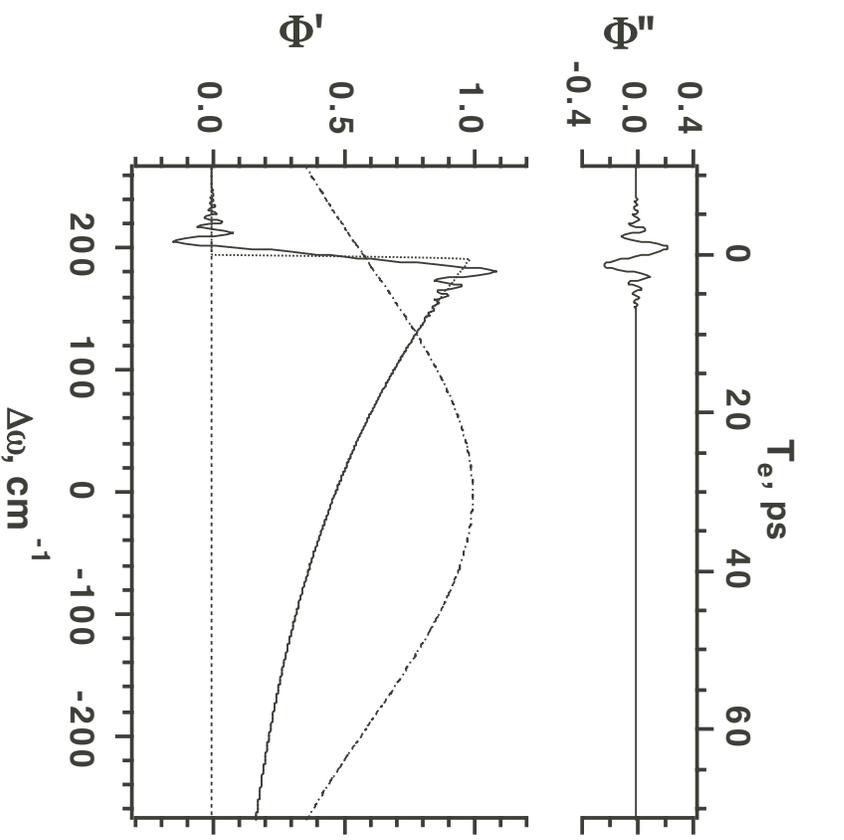

**Fig. 1**; Shkrob et al

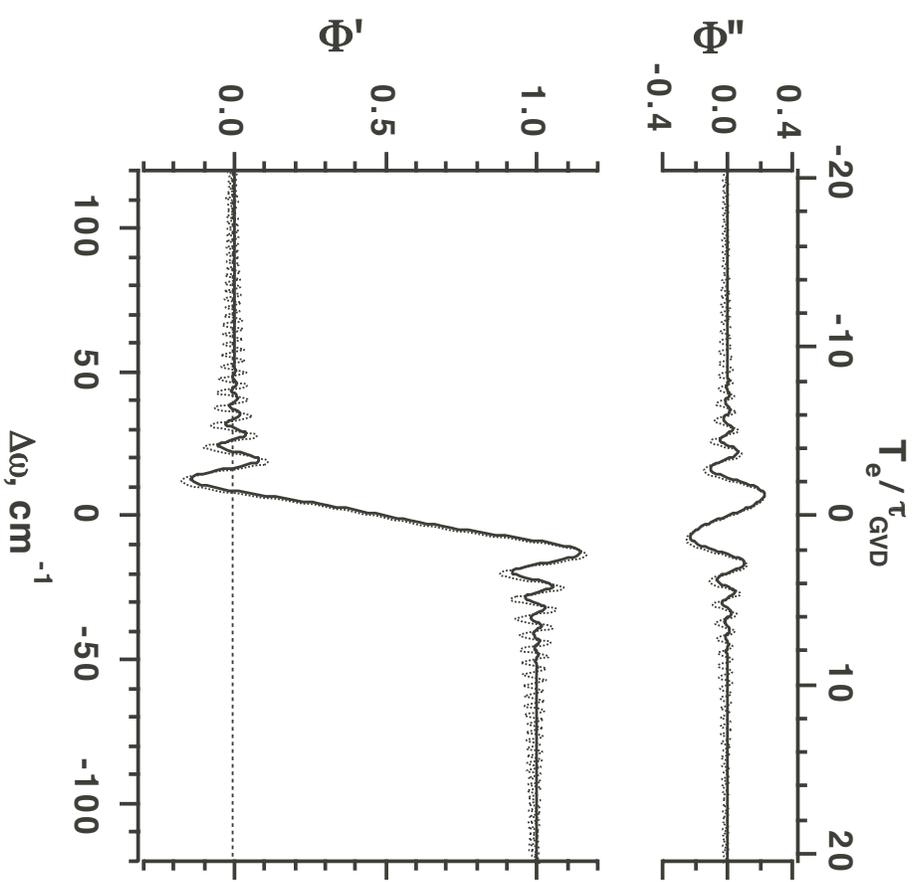

**Fig. 2**; Shkrob et al

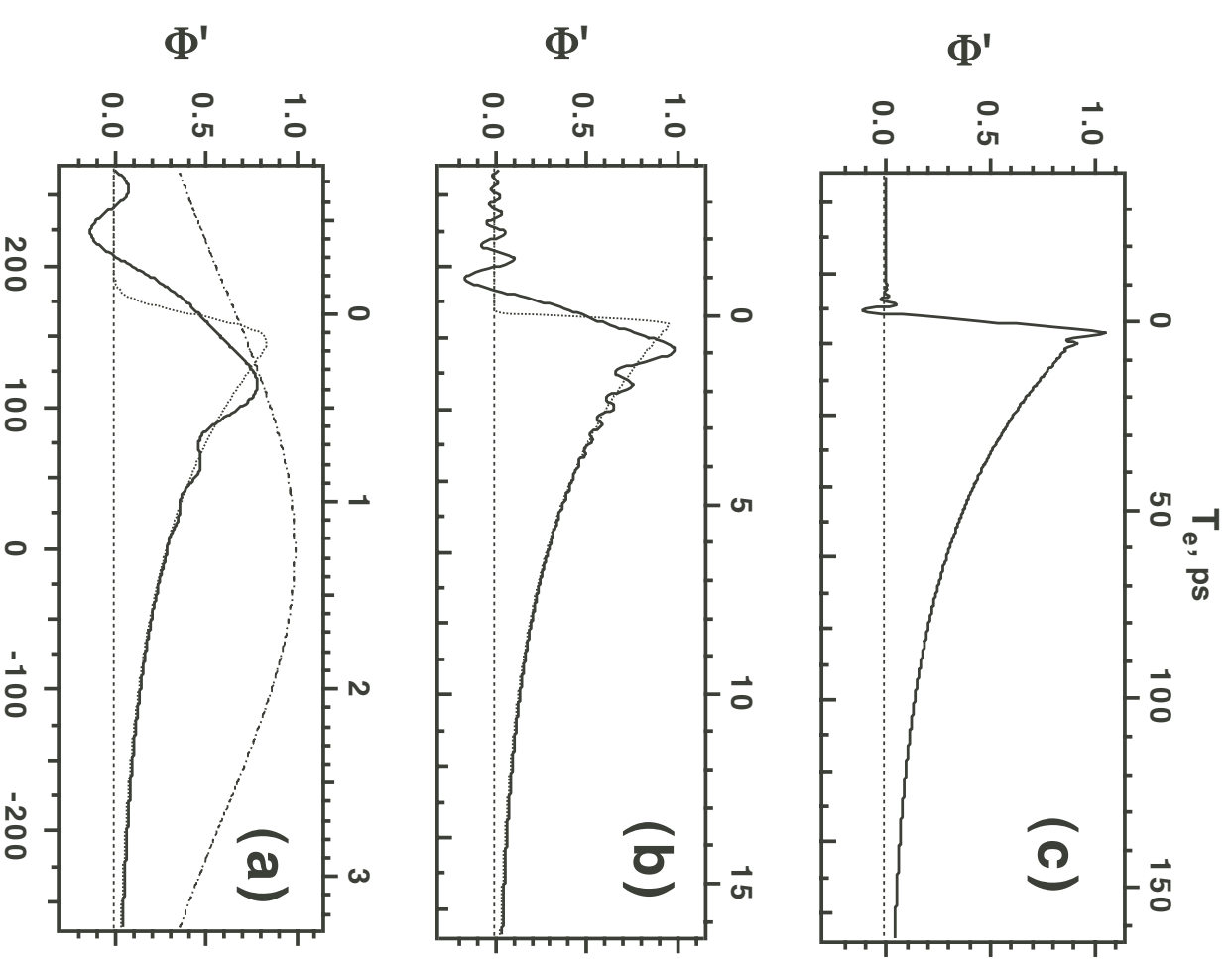

Fig. 3; Shkrob et al

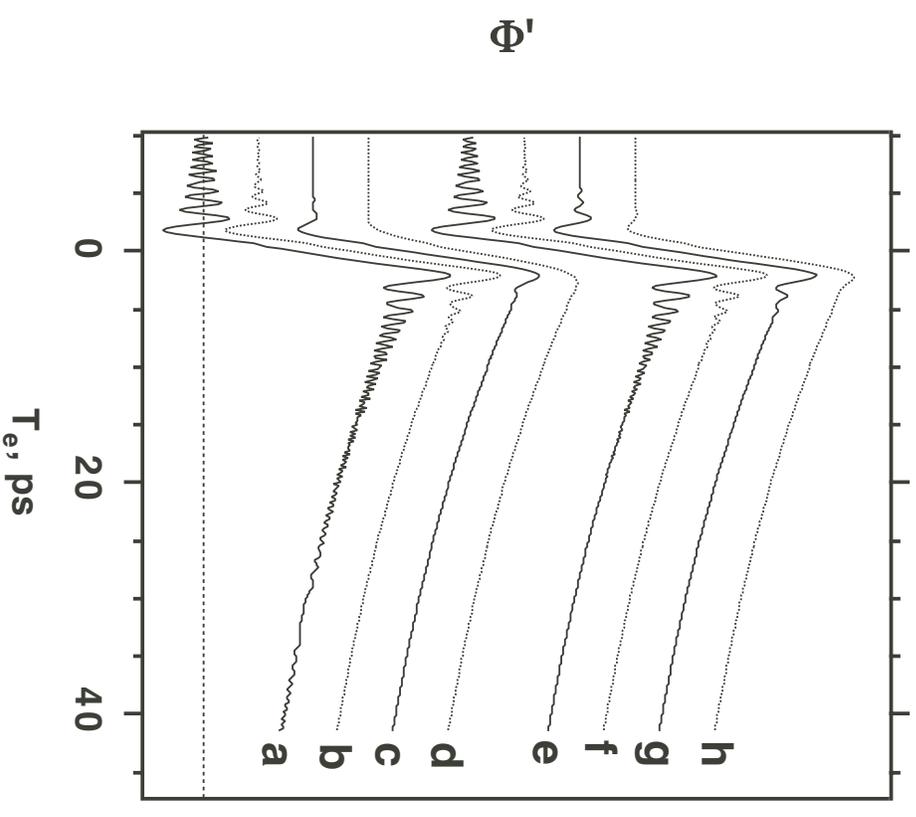

Fig. 4; Shkrob et al

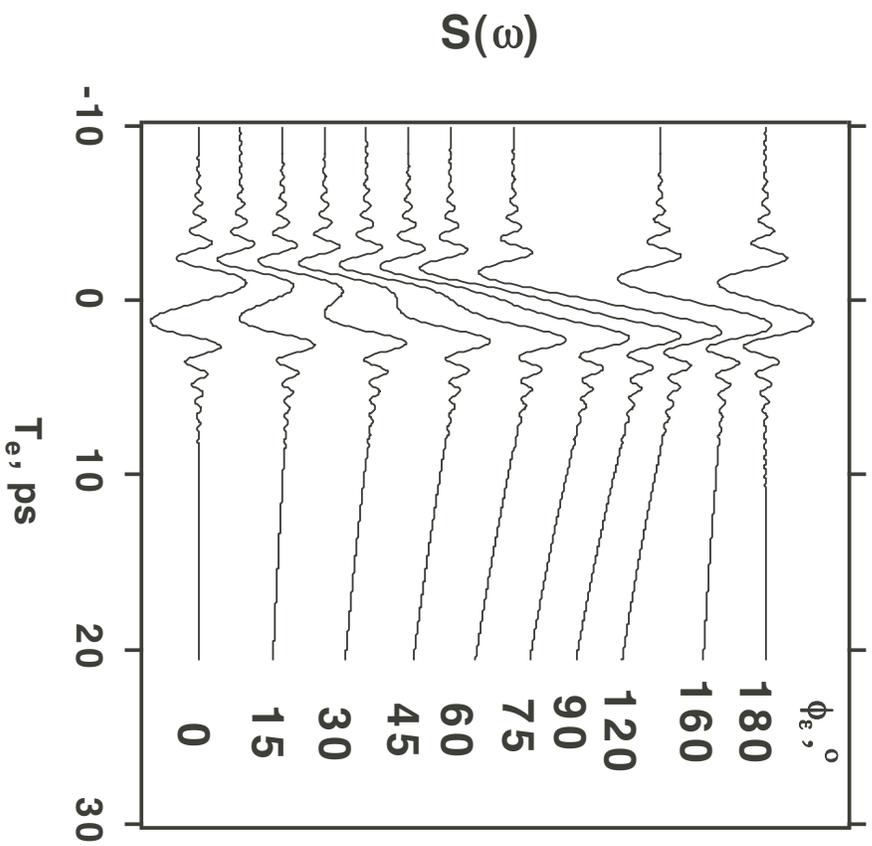

**Fig. 5**; Shkrob et al

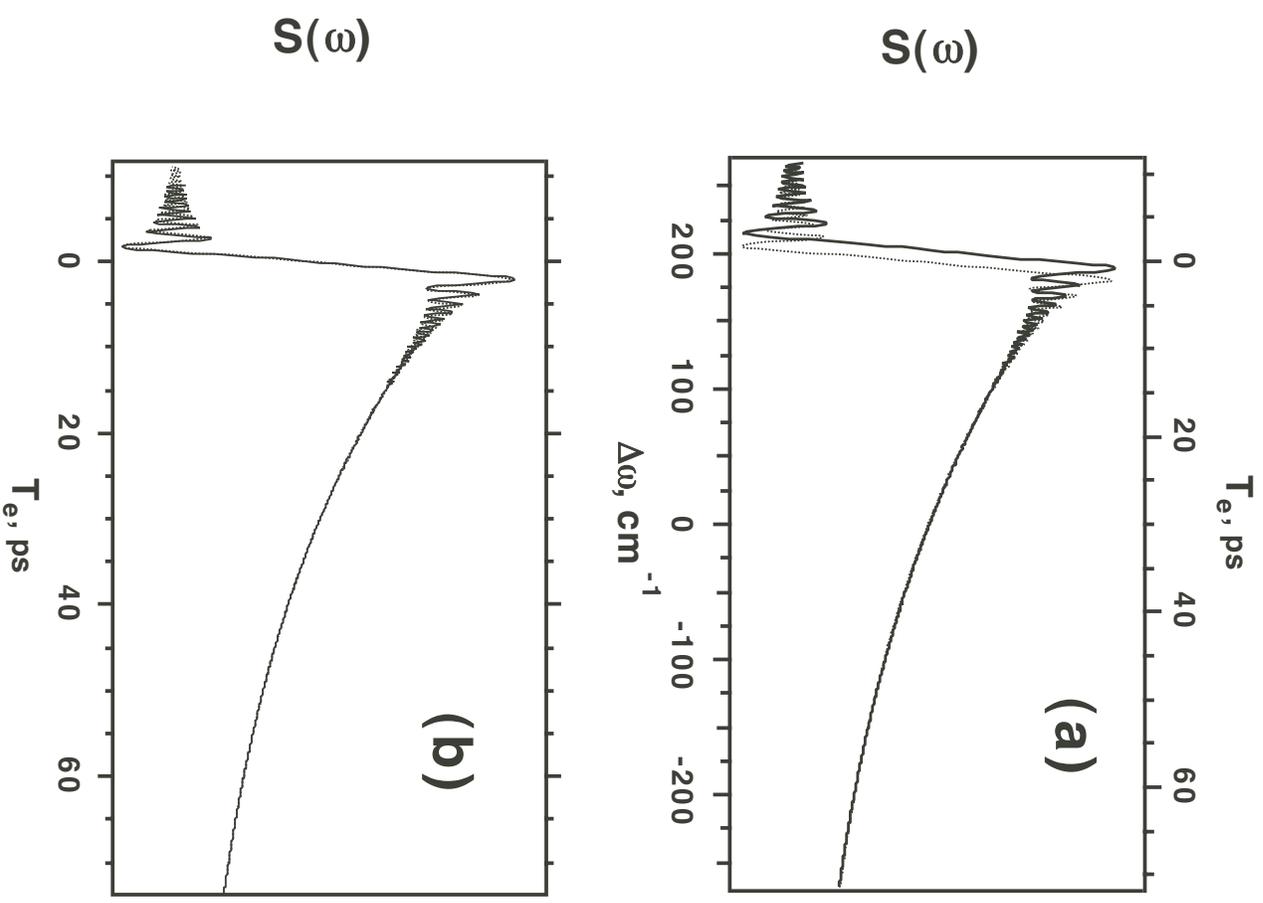

**Fig. 6**; Shkrob et al

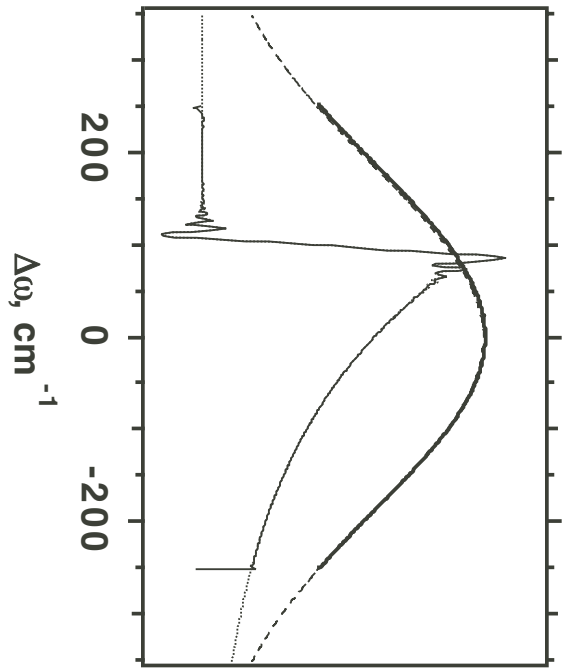

**Fig. 7; Shkrob et al**

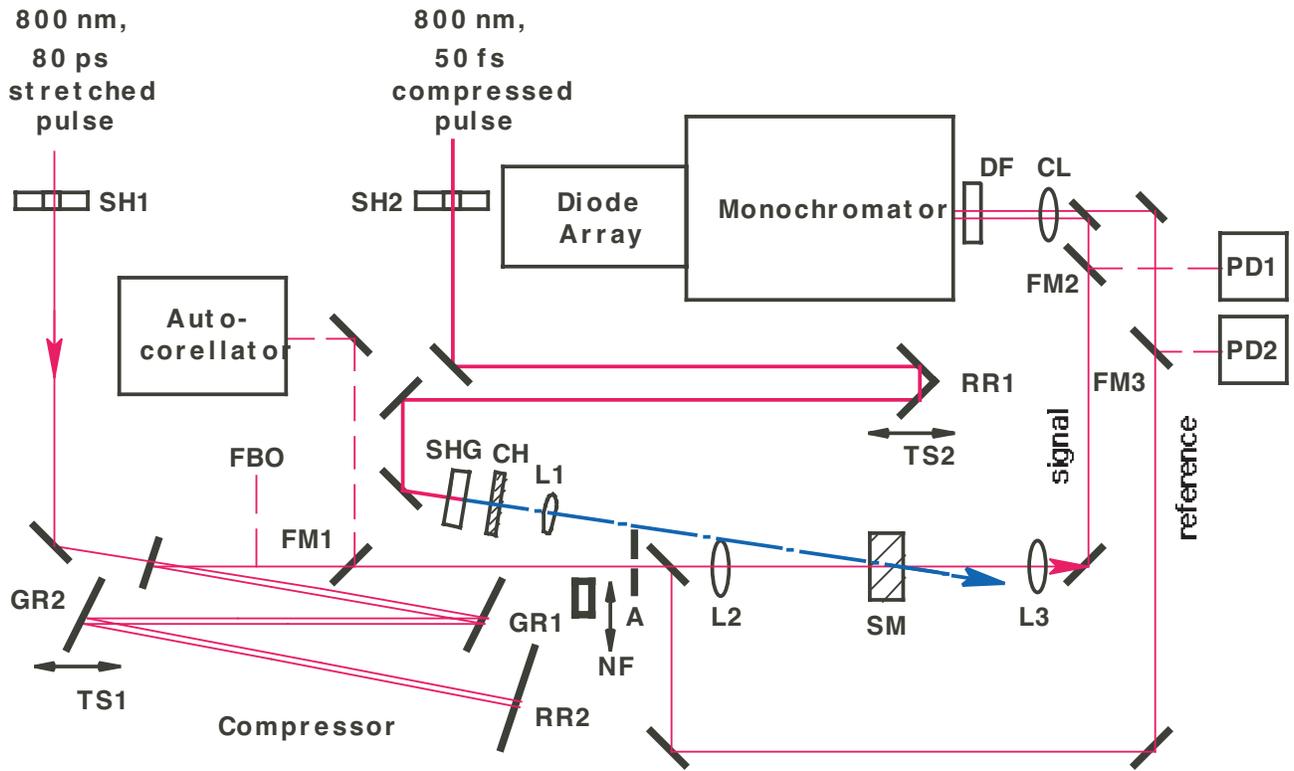

**Fig. 8; Shkrob et al**

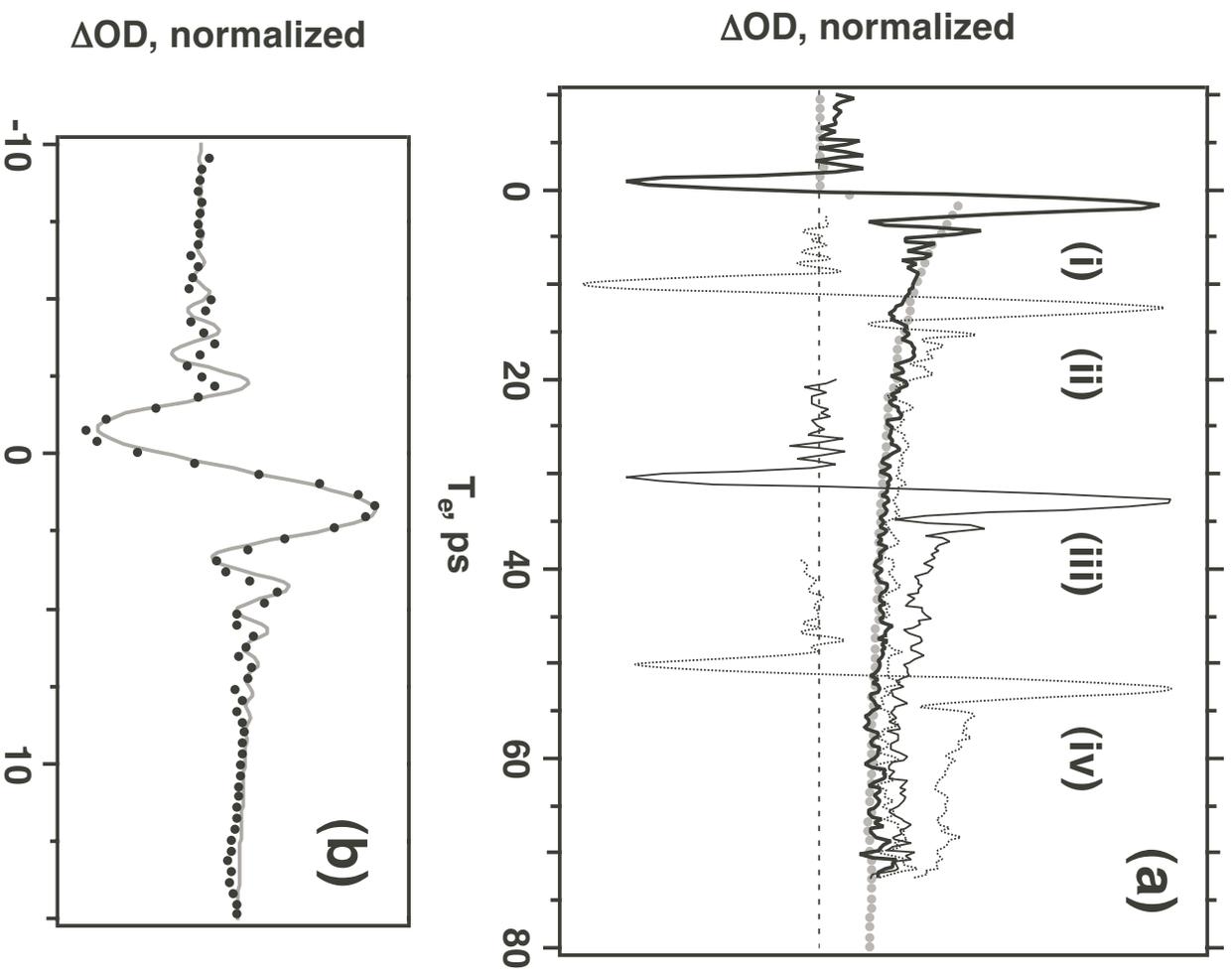

Fig. 9; Shkrob et al

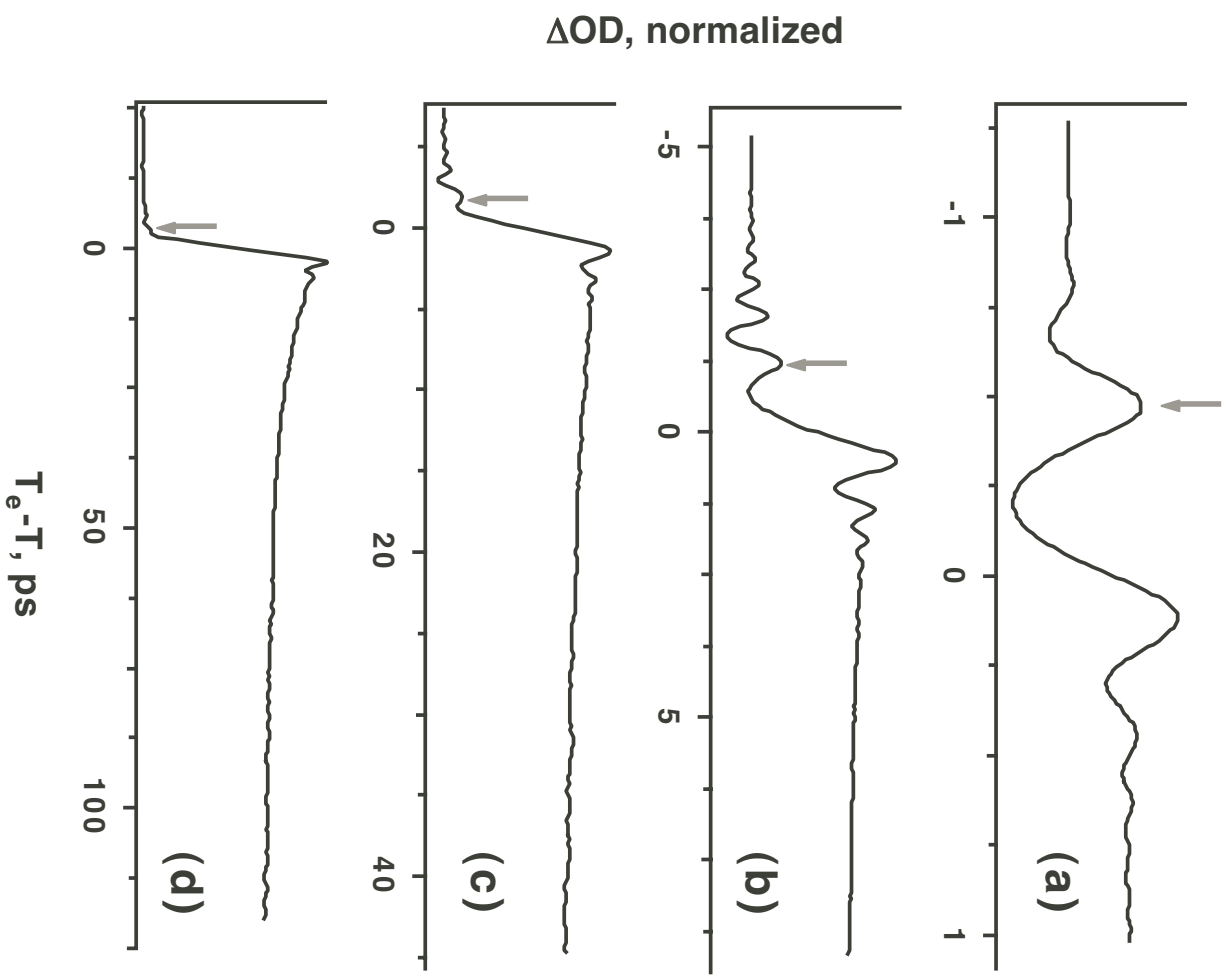

Fig. 10; Shkrob et al